\UseRawInputEncoding
\documentclass[superscriptaddress, twocolumn, amsmath, amssymb, aps,prl, letterdraft, notitlepage,longbibliography]{revtex4-1}
\usepackage{graphicx,graphics,epsfig,subfigure,times,bm,bbm,amssymb,amsmath,amsfonts,amsthm,mathrsfs,MnSymbol}
\usepackage[matrix,frame,arrow]{xypic}
\usepackage[normalem]{ulem}
\usepackage{slashed}
\usepackage{color}
\usepackage[usenames,dvipsnames,svgnames,table]{xcolor}
\usepackage[english]{babel}
\usepackage{verbatim}
\usepackage{tabularx}
\definecolor{darkblue}{rgb}{0.0,0.0,0.3}
\usepackage[colorlinks=true,
            linkcolor=red,
            urlcolor= darkblue,
            citecolor=blue]{hyperref}

\newcommand{\bea}{\begin{eqnarray}}
\newcommand{\eea}{\end{eqnarray}}

\begin{document}
\title{Periodically-driven quantum thermal machines from warming up to limit cycle}




\author{Junjie Liu}
\address{Department of Chemistry and Centre for Quantum Information and Quantum Control,
University of Toronto, 80 Saint George St., Toronto, Ontario, M5S 3H6, Canada}
\author{Kenneth A. Jung}
\address{Department of Chemistry, Stanford University, Stanford, California, 94305, USA}
\author{Dvira Segal}
\address{Department of Chemistry and Centre for Quantum Information and Quantum Control,
University of Toronto, 80 Saint George St., Toronto, Ontario, M5S 3H6, Canada}
\address{Department of Physics, 60 Saint George St., University of Toronto, Toronto, Ontario, M5S 1A7, Canada}

\begin{abstract}
%
%
Theoretical treatments of periodically-driven quantum thermal machines (PD-QTMs) 
are largely 
focused on the limit-cycle stage of operation 
characterized by a periodic state of the system. 
Yet, this regime is not immediately accessible for experimental verification. 
Here, we present a general thermodynamic framework that can handle
the performance of PD-QTMs both before and during the limit-cycle stage of operation. It is 
achieved by observing that periodicity  
may break down at the ensemble average level, even in the limit-cycle phase. 
With this observation, 
and using conventional thermodynamic expressions for work and heat,
we find that a complete description of the first law of thermodynamics for PD-QTMs requires 
a new contribution, 
which vanishes {\it only}  in the limit-cycle phase under rather weak system-bath couplings. Significantly, this contribution
is substantial at strong couplings {\it even at limit cycle}, thus largely affecting the behavior of the thermodynamic efficiency.
%
We demonstrate our framework by simulating a quantum Otto engine 
building upon a driven resonant level model. 
Our results provide new insights towards a complete description of 
 PD-QTMs, from turn-on to the limit-cycle stage and, particularly, shed light on the development of 
quantum thermodynamics at strong coupling.  
\end{abstract}

\date{\today}

\maketitle

{\it Introduction.--} 
A tantalizing goal of quantum thermodynamics is to investigate, analyze and ultimately design thermal machines 
on the quantum scale 
\cite{Scovil.59.PRL,Geva.92.JCP,Gemmer.09.NULL,Seifert.12.RPP,Pekola.15.NP,Kosloff.15.E,Goold.16.JPA,Anders.16.CP,Benenti.17.PR,Binder.18.NULL,Seifert.19.ARCMP}.
Among the various routes towards this vision, devising periodically-driven quantum thermal machines (PD-QTMs) 
in which work can be extracted through externally driving a quantum working substance represents a paradigmatic one.
This field has attracted much attention in recent years with intriguing theoretical proposals \cite{Brandner.15.PRX,Brandner.16.PRE,Proesmans.16.PRX,Newman.17.PRE,Alecce.15.NJP,Perarnau-Llobet.18.PRL,Mohammady.19.PRE,Kloc.19.PRE,Pezzutto.19.QST,Mohammady.19.PRE,ParkJ.19.PRE,Solfanelli.20.PRB,Altintas.20.QIP,Wiedmann.20.NJP,Shirai.21.PRR,Strasberg.21.PRL} 
together with delicate experimental realizations in various platforms 
\cite{Abah.12.PRL,Rossnagel.14.PRL,Dechant.15.PRL,Sood.16.NP,Robnagel.16.S,Peterson.19.PRL,Klatzow.19.PRL,Assis.19.PRL,Pekola.19.ARCMP,Lindenfels.19.PRL}.

Existing theories on PD-QTMs are largely carried out in the limit-cycle stage of operation,  
where the working substance should recoil to its original state after a driving cycle 
(see, e.g., Refs.\cite{Feldmann.04.PRE,Brandner.15.PRX,Brandner.16.PRE,Insinga.16.PRE,Proesmans.16.PRX,Insinga.18.PRE,Mohammady.19.PRE,Shirai.21.PRR,Menczel.19.JPA}). 
Although the limit-cycle regime enables theoretical simplicity, it faces potential challenges. 
On the one hand, recent studies have pointed out that the state of the working substance needs not inherit the 
periodicity of external drivings at finite times \cite{Menczel.19.JPA,Pezzutto.19.QST,Wiedmann.20.NJP}, 
namely, there may exist a transient warming-up phase for the operation of PD-QTMs before a limit cycle sets in. 
On the other hand, the limit-cycle stage is not immediately accessible experimentally, 
which may lead to discrepancy between theoretical predictions and experimental measurements. 
Addressing these challenges requires a unified thermodynamic description of both the warming-up and limit-cycle phases of PD-QTMs, which is, however, still missing.

In this work, we present a thermodynamic framework that deals with
cycle-number dependent thermodynamic quantities. As such, it 
allows us to characterize in a unified manner 
the performance of PD-QTMs in both the warming-up and limit-cycle phases. 
Our theory relies on the observation that periodicity can break down at the ensemble average level, namely,
$\langle\mathcal{O}(t)\rangle\neq\langle\mathcal{O}(t+\mathcal{T})\rangle$ with $\mathcal{O}$ being 
a periodic observable $\mathcal{O}(t)=\mathcal{O}(t+\mathcal{T})$ and $\mathcal{T}$ the driving period. 
This breakdown is trivial in the transient warming-up phase where even the state of the 
working substance does not behave in a periodic manner. However, nontrivially,
this breakdown can show up in the {\it limit-cycle} phase if $\mathcal{O}$ involves bath degrees of freedom, 
a scenario which is frequently encountered in studies of quantum thermodynamics 
(see recent discussions \cite{Talkner.20.RMP,Seshadri.21.PRB} and reference therein).

Building upon this observation, we find that the thermodynamic characterization of PD-QTMs generally acquires 
a new contribution in both the warming-up and limit-cycle phases,
besides the conventional notions of work and heat. 
Particularly, through completing the first law of 
thermodynamics [cf. Eq. (\ref{eq:first_law})] 
and ensuring its validity across the whole parameter space at all times, we identify this extra term (denoted as $\mathcal{A}$ hereafter) as the change of an ensemble average 
$\langle H_S(t)+H_I(t)\rangle$ within a cycle [cf. Eq. (\ref{eq:a_term})]; 
$H_{S(I)}$ denotes the Hamiltonian for the working substance (system-bath coupling). 
Intriguingly, this $\mathcal{A}$ term vanishes {\it only} in the limit-cycle phase provided that the system-bath coupling is rather weak. We further uncover that contributions from the $\mathcal{A}$ term play a nontrivial role in shaping behavior 
of the thermodynamic efficiency [cf. Eq. (\ref{eq:eta})] in both the warming-up and limit-cycle phases, especially at strong coupling, compared with existing characterizations, which largely neglect it.

\begin{figure}[tbh!]
\centering
\includegraphics[width=0.9\columnwidth]{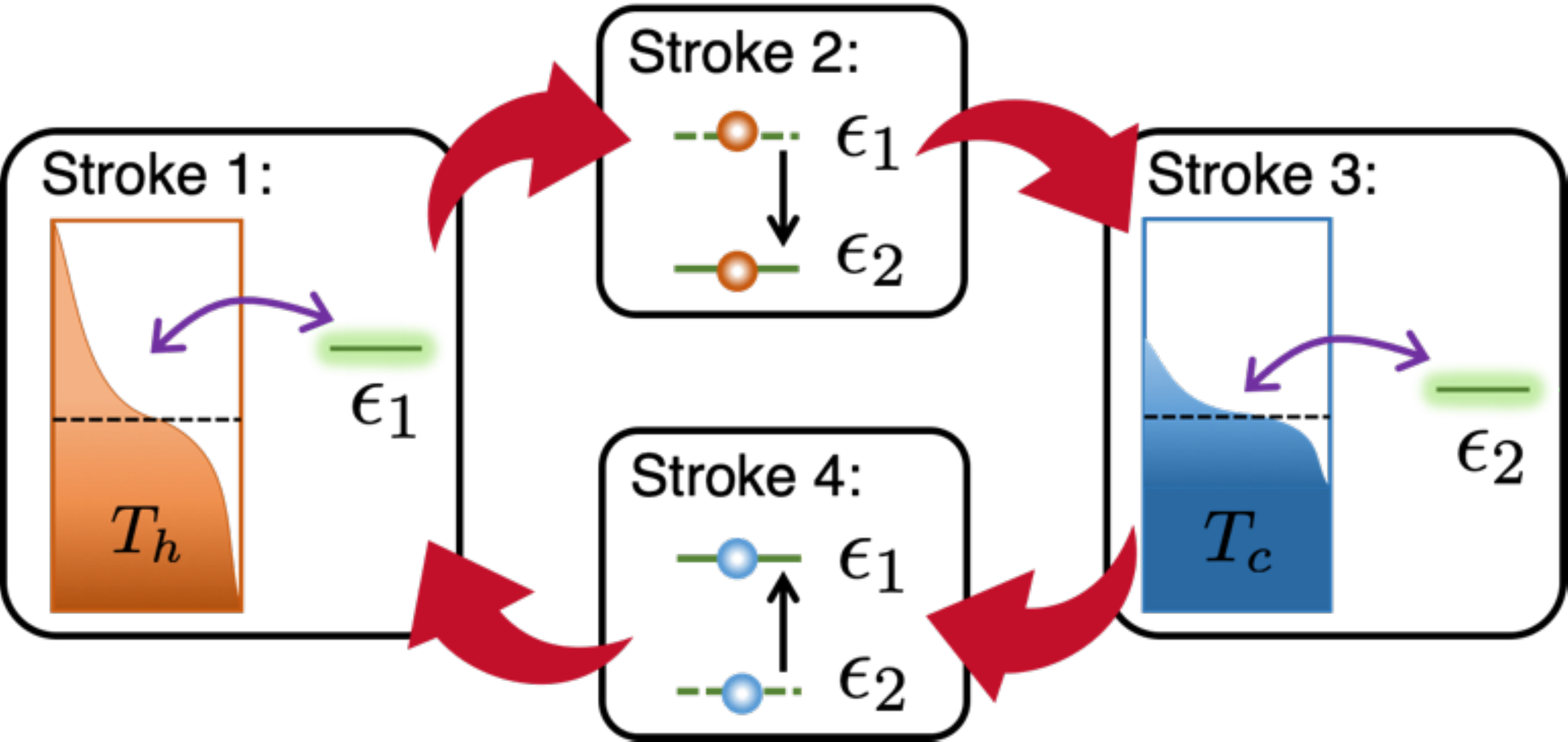} 
\caption{Schematic of a driven resonant level engine operating with a quantum Otto cycle. 
The cycle consists of isochoric heating (stroke 1) and cooling (stroke 3) processes 
and two work-extracting processes (strokes 2 and 4 with $\epsilon(t)$ tuned from $\epsilon_1$ to $\epsilon_2$ and from $\epsilon_2$ to $\epsilon_1$, respectively). 
}
\protect\label{fig:otto}
\end{figure}

\begin{figure*}[thb!]
 \centering
\includegraphics[width=2\columnwidth]{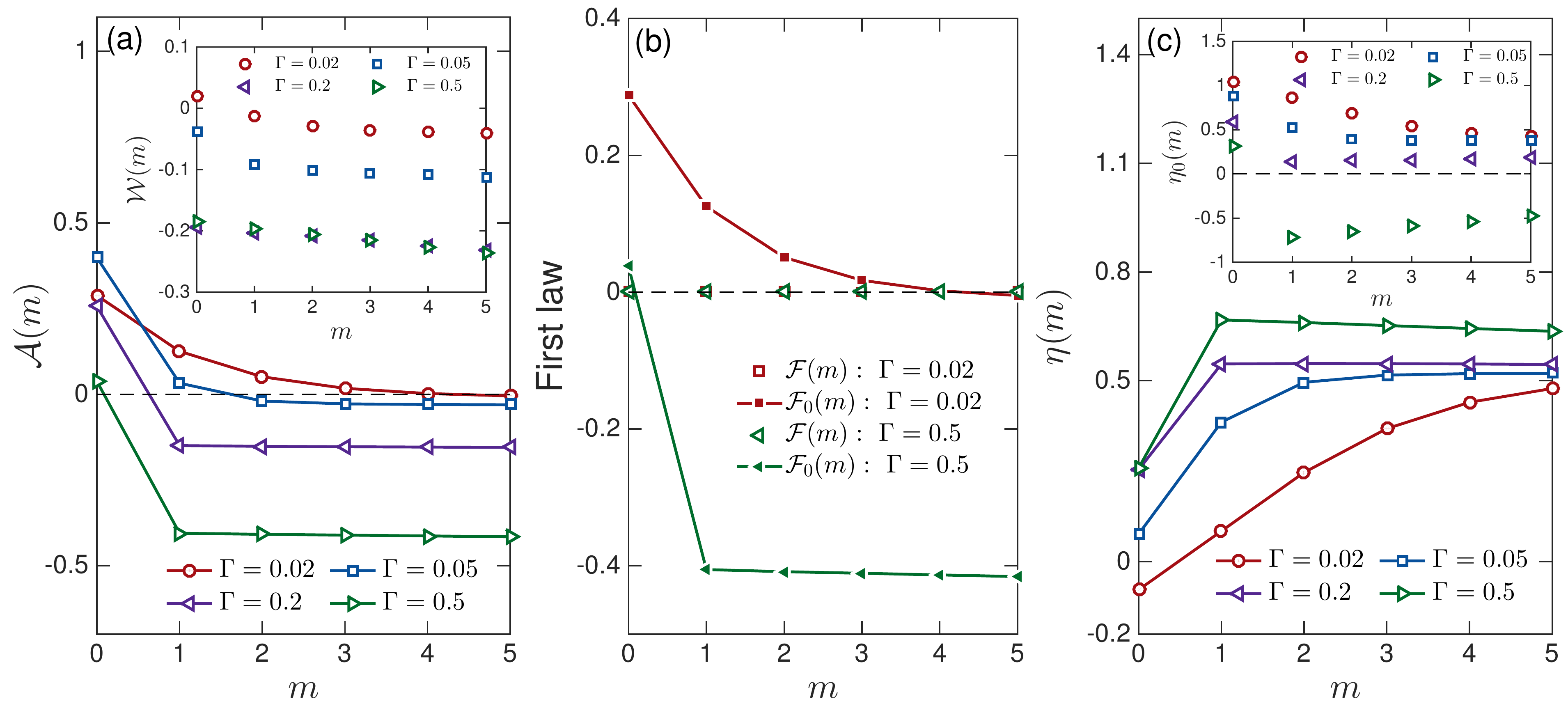} 
%
\caption{(a) 
$\mathcal{A}(m)$ [cf. Eq. (\ref{eq:a_term})] as a function of the number of cycles $m$ 
(note $m+1$ is the number of completed cycles) 
while varying the coupling strengths, 
$\Gamma=0.02$ (red circles), $\Gamma=0.05$ (blue squares), $\Gamma=0.2$ (purple left-triangles) and $\Gamma=0.5$ (green right-triangles). 
Inset: Net injected work $\mathcal{W}(m)$ 
[cf. Eq. (\ref{eq:work})] as a function of $m$ while varying coupling strengths.
(b) Comparison between 
the function $\mathcal{F}(m)\equiv\mathcal{W}(m)+\sum_v\mathcal{Q}_v(m)-\mathcal{A}(m)$ (empty symbols), which constitute the correct first law, and its incomplete counterpart $\mathcal{F}_0(m)\equiv\mathcal{W}(m)+\sum_v\mathcal{Q}_v(m)$ (solid symbols) as a function of $m$ for $\Gamma=0.02$ (red squares) and $\Gamma=0.5$ (green triangles).
 (c) Thermodynamic efficiency $\eta(m)$ defined for a cycle $[m\mathcal{T},(m+1)\mathcal{T}]$ 
[cf. Eq. (\ref{eq:eta})] as a function of $m$ with varying coupling strengths 
$\Gamma=0.02$ (red circles), $\Gamma=0.05$ (blue squares), $\Gamma=0.2$ (purple left-triangles) and $\Gamma=0.5$ (green right-triangles). 
Inset: The incomplete efficiency $\eta_0(m)=1+\mathcal{Q}_c(m)/\mathcal{Q}_h(m)$ 
as a function of $m$.
Lines in the main plots are drawn for guidance. 
Other dimensionless parameters are $\beta_h=0.2$, $\beta_c=1.5$, $\mu_h=\mu_c=0$, $\epsilon_1=2$, 
$\epsilon_2=1$, $t_1=t_3=\mathcal{T}/3$ and $t_2=t_4=\mathcal{T}/6$ with $\mathcal{T}=60$.
}
\protect\label{fig:eta}
\end{figure*}

We illustrate our general framework using the quantum Otto heat engine 
(see Fig. \ref{fig:otto} for a sketch), with a driven resonant electronic level as the working substance.
Based on simulations, we confirm our theoretical predictions of
the behaviors of thermodynamic quantities incorporating the $\mathcal{A}$ term, 
including the first law of thermodynamics and the efficiency  (see Fig. \ref{fig:eta}).
In comparison, conventional definitions without $\mathcal{A}$ fail to capture the correct thermodynamics, 
especially at strong coupling where the $\mathcal{A}$ term is substantial in both warm-up and limit-cycle regimes.
The framework presented here allows the description of PD-QTMs even before the limit cycle stage, 
and at strong coupling, both aspects crucial for describing experimental systems and advancing
the field of quantum thermodynamics.


{\it Quantum thermodynamics of generic PD-QTMs.--} 
Generally, PD-QTMs can be modeled as ($\hbar=1$, $k_B=1$ hereafter)
\begin{equation}\label{eq:H}
H(t)~=~H_S(t)+H_B+H_I(t). 
\end{equation}
Here, $H_S(t)$ denotes a semiclassically-driven working substance with the understanding that the working substance exchanges work with an external 
agent which we do not explicitly include within the quantum description. $H_B=\sum_{v=h,c}H_B^v$ consists of a hot (h) and a cold (c) thermal bath and $H_I(t)=\sum_vH_I^v(t)\equiv\sum_v\lambda_v(t)H_I^v$ denotes a time-dependent system-bath coupling. Here, $\lambda_v(t)$ is a time-dependent coefficient describing the contact switching of the $v$ reservoir, used to implement thermodynamic strokes (see, e.g., Refs. \cite{Wiedmann.20.NJP,Pancotti.20.PRX,Shirai.21.PRR}). We assume periodic protocols $H_{S,I}(t)=H_{S,I}(t+\mathcal{T})$ such that $H(t)=H(t+\mathcal{T})$ with $\mathcal{T}$ the period of the driving.

To analyze the quantum thermodynamics of the PD-QTM, 
we introduce the definitions of heat and work based on the complete Hamiltonian dynamics, $H(t)$ \cite{Bruch.16.PRB,Oz.20.JCTC,Wiedmann.20.NJP,Cangemi.21.PRR}. This approach naturally avoids the problem of the partitioning of the system-bath interaction.
Denoting by $\rho_{\mathrm{tot}}$ the total density matrix of the composite system, 
the integrated injected work $\mathcal{W}(m)$ and absorbed heat $\mathcal{Q}_{v}(m)$ from the $v$ bath during the $m$-th 
cycle $[m\mathcal{T}, (m+1)\mathcal{T}]$ with a non-negative integer $m\ge0$ are uniquely defined as \cite{Bruch.16.PRB,Oz.20.JCTC,Wiedmann.20.NJP,Cangemi.21.PRR}
\bea
\mathcal{W}(m) &\equiv& \int_{m\mathcal{T}}^{(m+1)\mathcal{T}}\,dt\mathrm{Tr}\left[\dot{ H}(t)\rho_{\mathrm{tot}}(t)\right],\label{eq:work}\\
\mathcal{Q}_{v}(m) &\equiv& -\int_{m\mathcal{T}}^{(m+1)\mathcal{T}}\,dt\mathrm{Tr}\left[H_B^{v}\dot{\rho}_{\mathrm{tot}}(t)\right].\label{eq:heat}
\eea
Here, $\dot{\mathcal{X}}(t)\equiv d\mathcal{X}(t)/dt$ for an arbitrary operator $\mathcal{X}(t)$, the trace is performed over system and bath degrees of freedom. 
The introduction of the cycle-number ($m$) dependent thermodynamic quantities allows us to characterize 
the performance of PD-QTMs in both the warming-up and limit-cycle stages. 
Note that the expression for absorbed heat $\mathcal{Q}_{v}$ is consistent with the definition employed
in quantum transport scenarios \cite{Esposito.15.PRL,Esposito.15.PRB,Bergmann.21.EPJ}. 
For the work definition, it naturally incorporates work costs for implementing a time-dependent system-bath coupling \cite{Newman.17.PRE}. In our convention, a heat engine corresponds to $\mathcal{W}<0$, $\mathcal{Q}_h>0$ and $\mathcal{Q}_c<0$. 

Since the integrand in Eq. (\ref{eq:work}) is just $d\langle H(t)\rangle/dt$ 
with $\langle H(t)\rangle\equiv\mathrm{Tr}[H(t)\rho_{\mathrm{tot}}(t)]$ 
the internal energy of the composite system, we can express Eq. (\ref{eq:work}) as 
$\mathcal{W}(m)=\mathrm{Tr}\big\{H(t)[\rho_{\mathrm{tot}}((m+1)\mathcal{T})-\rho_{\mathrm{tot}}(m\mathcal{T})]\big\}$. 
This immediately leads to an intriguing observation: 
$\rho_{\mathrm{tot}}(m\mathcal{T})\neq\rho_{\mathrm{tot}}((m+1)\mathcal{T})$, 
generally valid since $\mathcal{W}(m)\neq 0$ for a PD-QTM. 
Note that this difference does not prevent the existence of a limit-cycle phase for
 the {\it reduced system density matrix} $\rho_S$,
$\rho_S(m\mathcal{T})=\rho_S((m+1)\mathcal{T})$ 
\cite{Feldmann.04.PRE,Brandner.15.PRX,Brandner.16.PRE,Insinga.16.PRE,Proesmans.16.PRX,Insinga.18.PRE,Mohammady.19.PRE,Shirai.21.PRR,Menczel.19.JPA}, since the bath state needs not be a periodic function at all. 
This simple observation has a profound consequence: An operator $\mathcal{O}$ involving bath degrees of freedom 
generally shows $\langle\mathcal{O}(m\mathcal{T})\rangle\neq\langle\mathcal{O}((m+1)\mathcal{T})\rangle$,
even when $\mathcal{O}(m\mathcal{T})=\mathcal{O}((m+1)\mathcal{T})$. 
Namely, the periodicity can break down at the ensemble average level. 

We now turn to the first law of thermodynamics for PD-QTMs. In terms of our definitions
for heat and work, it takes the following exact form 
\footnote{One can obtain the present first law by combining the expression $\mathcal{W}(m)=\int_{m\mathcal{T}}^{(m+1)\mathcal{T}}dt\frac{d \langle H(t)\rangle}{dt}$ with the Hamiltonian decomposition of Eq. (\ref{eq:H}).},
\begin{equation}\label{eq:first_law}
\mathcal{W}(m)+\sum_{v}\mathcal{Q}_{v}(m)-\mathcal{A}(m)~=~0.
\end{equation}
Crucially, besides the conventional notions of work and heat, an extra quantity appears, $\mathcal{A}(m)$,
\begin{equation}\label{eq:a_term}
\mathcal{A}(m) 
\equiv \mathrm{Tr}\Big\{H_{SI}(m\mathcal{T})[\rho_{\mathrm{tot}}((m+1)\mathcal{T})-\rho_{\mathrm{tot}}(m\mathcal{T})]\Big\}.
\end{equation}
Here, we denote by $H_{SI}(m\mathcal{T})\equiv H_S(m\mathcal{T})+H_I(m\mathcal{T})$. 
In arriving at the first law Eq. (\ref{eq:first_law}), we emphasize that no approximations were invoked. 
The $\mathcal{A}$ term reflects the change to  ${\rm Tr} \left\{ H_{SI}\, \rho_{{\rm tot}} \right\}$
after a cycle, 
highlighting the nontrivial role of the system-bath coupling in quantum engines \cite{Shirai.21.PRR,Talkner.20.RMP}. 
In the warming-up phase where even the system density matrix $\rho_S$ does not inherit the periodicity of the driving 
protocol, one naturally expects a nonzero $\mathcal{A}(m)$. 
In the limit-cycle phase, $\mathcal{A}(m)$ vanishes {\it only} when the system-bath coupling is negligible 
compared to the system energy scale since then 
$\mathcal{A}(m)\simeq\mathrm{Tr}_S\Big\{H_S(m\mathcal{T})[\rho_{S}((m+1)\mathcal{T})-\rho_{S}(m\mathcal{T})]\Big\}=0$. Otherwise, at strong coupling, we still have $\mathcal{A}(m)\neq 0$ 
as $\langle H_{I}(m\mathcal{T})\rangle\neq\langle H_{I}((m+1)\mathcal{T})\rangle$
 [see discussion before Eq. (\ref{eq:first_law})].
This result contradicts previous studies (see, e.g., Ref. \cite{Wiedmann.20.NJP}), 
which neglected the contribution from $\mathcal{A}(m)$ regardless of system-bath coupling strength. 
%
This first law and the nontrivial behavior of $\mathcal{A}(m)$ represent our first main result. As for the second law of thermodynamics, we still observe the conventional Clausius inequality \cite{Esposito.10.NJP,Brandner.16.PRE,Seshadri.21.PRB}, $\mathcal{S}(m)\equiv[\mathcal{S}_S((m+1)\mathcal{T})-\mathcal{S}_S(m\mathcal{T})]-\sum_v\beta_v\mathcal{Q}_v(m)\ge0$ with $\mathcal{S}_S(m)=-\mathrm{Tr}[\rho_S(m\mathcal{T})\ln \rho_S(m\mathcal{T})]$ the von Neumann entropy of the working substance. 

%
%
Using Eq. (\ref{eq:first_law}), the averaged thermodynamic efficiency over a cycle, 
$\eta(m)\equiv-\frac{\mathcal{W}(m)}{\mathcal{Q}_{h}(m)}$, reads \bea\label{eq:eta}
\eta(m)~=~\eta_0(m)-\frac{\mathcal{A}(m)}{\mathcal{Q}_{h}(m)}.
\eea
Here, $\eta_0=1+\mathcal{Q}_{c}/\mathcal{Q}_{h}$ 
is the common definition of efficiency 
\cite{Alecce.15.NJP,Mohammady.19.PRE,ParkJ.19.PRE,Solfanelli.20.PRB,Altintas.20.QIP},
which has been applied to strong coupling scenarios \cite{Perarnau-Llobet.18.PRL,Wiedmann.20.NJP}.  
Intriguingly, $\eta(m)$ gains an additional contribution $\mathcal{A}(m)/\mathcal{Q}_{h}(m)$, 
compared to $\eta_0(m)$. 
we point out that Eq. (\ref{eq:eta}) is completely general as no assumptions were invoked in its derivation: 
It can be applied to arbitrary PD-QTMs in both the warming-up and limit-cycle phases, 
regardless of the system-bath coupling strength. 
On the contrary, $\eta_0(m)$ can only be applied to the weak-coupling scenario in the 
limit-cycle phase where $\mathcal{A}(m)$ vanishes [see discussion below Eq. (\ref{eq:a_term})].
This working definition for the efficiency is our second main result.

{\it Example: Driven resonant level engine.--} 
To substantiate and exemplify our main results 
[cf. Eqs. (\ref{eq:first_law}), (\ref{eq:a_term}) and (\ref{eq:eta})] 
we present a numerical example of a quantum Otto heat engine 
(see Fig. \ref{fig:otto}) that extracts work from heat baths with a resonant level model 
\cite{Ludovico.14.PRB,Esposito.15.PRL,Esposito.15.PRB,Ochoa.16.PRB,Bruch.16.PRB,Thingna.17.PRE,Dou.18.PRB,Oz.19.MP,Dou.20.PRB,Oz.20.JCTC}, 
\bea\label{eq:resonant}
H_S(t) &=&\epsilon(t)c_d^{\dagger}c_d,\nonumber\\
H_B^v &=& \sum_k\epsilon_{kv}c_{kv}^{\dagger}c_{kv},\nonumber\\
H_I^v(t) &=& \lambda_v(t)\sum_k t_{kv}(c_{kv}^{\dagger}c_d+c_d^{\dagger}c_{kv}).
\eea
Here, the working substance consists of a single spinless electronic level with the annihilation operator 
$c_d$ and time-dependent energy $\epsilon(t)$. The dot is alternating its coupling to two fermionic baths (leads) 
$v=h,c$ of different temperatures.
$c_{kv}$ annihilates an electron with energy $\epsilon_{kv}$ in the $v$-lead that 
couples to the central level with $t_{kv}$ as the tunneling rate. 
The $v$-lead is occupied according to the Fermi-Dirac distribution with an inverse temperature $\beta_v$ 
and a chemical potential $\mu_v$; for our purpose, we set $\beta_h\neq\beta_c$ and $\mu_{h,c}=0$. 
The coupling of the central level to the leads is given by the
spectral densities $\Gamma_v(\epsilon)=2\pi\sum_kt_{kv}^2\delta(\epsilon-\epsilon_{kv})$. 
In what follows, we consider the wideband limit such that $\Gamma_v(\epsilon)=\Gamma_v$ 
and adopt symmetric couplings, $\Gamma_{h,c}=\Gamma$. 
The time-dependent coefficients $\lambda_v(t)$ equal $1$ when the coupling to the $v$-lead is turned on and $0$ otherwise, thus enabling thermalization strokes in the Otto cycle. 

The complete Otto cycle with period $\mathcal{T}$ consists of four strokes; see Fig. \ref{fig:otto} 
for an illustration: 
(1) Stroke 1 ($0\le t<t_1$): In this isochoric thermalization step, 
the central level with a fixed energy $\epsilon_1$ is coupled to the hot lead. 
(2) Stroke 2 ($t_1\le t<t_1+t_2$): The energy of the isolated central level is shifted from $\epsilon_1$ to $\epsilon_2$. 
(3) Stroke 3 ($t_1+t_2\le t<t_1+t_2+t_3$): In this isochoric thermalization step, the central level with a fixed energy $\epsilon_2$ is coupled to the cold lead. 
(4) Stroke 4 ($t_1+t_2+t_3\le t<\mathcal{T}$): The energy of the isolated central level is tuned back from $\epsilon_2$ to $\epsilon_1$. 
We remark that the adopted cycle protocol does not enforce a periodicity on the system density matrix {\it a priori} 
as we just tune the energy of the central level. 
The engine needs a transient warming-up process before settling into a limit-cycle operation mode. 
Moreover, we do not enforce an adiabatic work-extracting process, 
namely, strokes 2 and 4 can span finite times. In fact, the amount of work generated during strokes 2 and 4 is
independent of the detailed functional form of $\epsilon(t)$ connecting fixed end points $\epsilon_{1,2}$ 
as the charge occupation of the central level during these two strokes remains fixed; 
in Ref. \cite{SM}, we exploit this fact together with an equilibrium charge occupation expression \cite{LiuJ.20.NL} (note that we attach the central level to two leads alternatively in the Otto cycle) 
to estimate the heat-engine operation regime and determine relevant values of $\epsilon_{1,2}$ to be used in simulations.

The cycle number-dependent thermodynamics is simulated by employing the driven Liouville von-Neumann 
equation method \cite{Ajisaka.12.PRB,Zelovich.14.JCTC,Chen.14.JPCC,Zelovich.17.JCP,Chiang.20.JCP} 
whose performance for addressing quantum thermodynamics up to strong coupling has been carefully assessed 
\cite{Oz.19.MP,Oz.20.JCTC}; we relegate simulation details, including numerical implementation, initial condition at $t=0$ and adopted parameter values, to  \cite{SM}. 
A typical set of thermodynamic results is depicted in Fig. \ref{fig:eta}. Quantities at $m=0$ are evaluated for the first cycle $[0,\mathcal{T}]$.  

The behavior of $\mathcal{A}(m)$ along with the work $\mathcal{W}(m)$ is presented in Fig. \ref{fig:eta} (a); 
results for the heat exchange can be found in \cite{SM}. 
We find that as the coupling strength increases, the magnitude of $\mathcal{A}(m)$ becomes more pronounced.
From this it is evident that based on our heat and work definitions, 
the complete first law of thermodynamics {\it at all times} 
should read $\mathcal{W}+\sum_v\mathcal{Q}_v-\mathcal{A}=0$ as depicted in Fig. \ref{fig:eta} (b). 
In contrast, the sum $\mathcal{W}+\sum_v\mathcal{Q}_v$ considered in other 
studies (see, e.g., Refs. \cite{Mohammady.19.PRE,Wiedmann.20.NJP}) approaches zero only in the weak-coupling scenario,
and in the limit-cycle stage (see red line for $\Gamma=0.02$ in Fig. \ref{fig:eta} (b) at $m=5$ where the engine  
is reaching the limit-cycle phase; this is confirmed in \cite{SM} with a long-time simulation for $\Gamma=0.02$).
Importantly, beyond the weak coupling regime, the faulty first law that neglects the $\mathcal{A}$ term deviates from zero {\it at all times}.
Perhaps most surprisingly, Fig. \ref{fig:eta} (c) reveals that if one adopts the 
incomplete 
definition of efficiency, i.e. Eq. (\ref{eq:eta}) without $\mathcal{A}$, 
at strong coupling one would predict no heat engine behavior in this regime (see inset of Fig. \ref{fig:eta} (c)). 
However, by properly including the $\mathcal{A}$ contribution, thus accounting for the effect of system-bath coupling,
we find that the system does operate as a heat engine, consistent with the negative injected work value 
reported in the inset of Fig. \ref{fig:eta} (a). 

Besides the aforementioned basic features that reinforce the validity of our general theory, 
there are a few points worth discussing further: 
(i) Weak coupling scenarios need more transient cycles to warm up than the 
strong coupling counterparts before a limit-cycle phase sets in. 
Comparing the behaviors of $\mathcal{A}(m)$ and $\eta(m)$ as a function of $m$, 
we argue that the former can assess the number of transient cycles based on its magnitude variation 
with increasing $m$. 
For instance, from Fig. \ref{fig:eta} (a), we observe that an engine of $\Gamma=0.05$ requires at 
least 4 transient cycles; note $m+1$ is the number of completed cycles.
In contrast, an engine with $\Gamma=0.5$ needs just one cycle to warm up. We have verified that the number of transient cycles is independent of the initial charge occupation 
of the central level. 
(ii) The thermodynamic efficiency $\eta(m)$ does not reach a stationary value at strong couplings with increasing $m$  
(see curve for $\Gamma=0.5$ in Fig. \ref{fig:eta} (c)), 
%
noting that the heat $\mathcal{Q}_{h,c}(m)$ depends on the state of the 
bath, which needs not be a periodic function; see \cite{SM} for the behavior of $\mathcal{Q}_{h,c}(m)$ 
while varying $m$ and the coupling strength. This finding implies that a characterization based on a {\it single} cycle in the limit-cycle phase may not be applicable at strong couplings. 
(iii) In our simulations, $\eta(m)$ increases as the coupling strength increases. This seems at odds with some studies 
\cite{Perarnau-Llobet.18.PRL,Newman.20.PRE}, which have observed the opposite trend. 
However, we point out that these studies used the incomplete definition $\eta_0$. 
Indeed, the inset of Fig. \ref{fig:eta} (c)
shows that $\eta_0$ {\it decreases} as $\Gamma$ increases, until it becomes invalid for very large coupling strength. 
This suggests that interactions with the baths are in fact helpful in the heat-work conversion process \cite{Shirai.21.PRR}. The dramatic qualitative difference between the conventional definition $\eta_0$ and the correct
expression $\eta$ demonstrates why it is crucial to adopt the complete expressions, including the 
 $\mathcal{A}$ term. 

{\it Discussion.--} 
One may argue that the $\mathcal{A}$ term [cf. Eq. (\ref{eq:a_term})] can be absorbed by redefining 
heat and work (see discussions in Refs. \cite{Bruch.16.PRB,Ochoa.16.PRB,Dou.18.PRB}). 
We would like to point out that (i) the work definition we adopted [cf. Eq. (\ref{eq:work})] 
already comprises the internal energy change of the total composite system, 
and (ii) introducing an effective bath Hamiltonian with the replacement $H_B^v\to H_B^v+H_I^v/2$ \cite{Bruch.16.PRB,Ochoa.16.PRB,Dou.18.PRB} and accordingly modifying the heat definition [cf. Eq. (\ref{eq:heat})], 
one is still not able to eliminate the $\mathcal{A}$ term,
 by comparing the definitions Eq. (\ref{eq:heat}) and (\ref{eq:a_term}). 
After all, as we argued, the $\mathcal{A}$ term originates from the fact that periodicity breaks down at the ensemble average level even in the limit-cycle phase.

Recognizing the generic presence of $\mathcal{A}$ , it is tempting to explore whether 
the inclusion of work costs for the attachment and detachment of the working substance to and from the baths 
can affect the behavior of $\mathcal{A}$.
While here we switched between strokes by tuning  $\lambda_v(t)$ [see Eq. (\ref{eq:resonant})] 
using a sharp step function alternating between  $0$ and $1$, one could adopt a smooth function.
We leave this point to future investigations.

Although our thermodynamic framework is applicable to general PD-QTMs, 
 a direct measurement of $\mathcal{A}$ is experimentally challenging since 
it requires measuring energy change associated with the system-bath coupling. 
However, the utility of our framework is that it provides predictions for efficiency 
as a function of the cycle number, enabling a direct comparison with experiments. 
Exploring feasible measurement schemes for quantifying the role of $\mathcal{A}$, 
in light of recent experimental advances in single-electron and superconducting circuits 
\cite{Pekola.15.NP,Pekola.19.ARCMP}, is an interesting future direction.

In summary, we have presented a unified, consistent thermodynamic theory, with cycle-number dependent thermodynamic quantities, 
which describes the operation of PD-QTMs both before and during the limit-cycle phase, 
covering strongly-coupled thermal machines.
The unearthed $\mathcal{A}$ term (i) provides a complete description of the first law of thermodynamics and (ii)
substantially modifies the efficiency of the engine, particularly at strong coupling, 
compared with existing characterizations. 
With its ability to describe PT-QTMs from the warming up phase to their limit cycle 
and from weak to strong coupling,
we expect this framework to promote the experimental development of quantum thermal machines.

{\it Acknowledgement.--}
J.L. and D.S. acknowledge support from the Natural Sciences and Engineering Research Council (NSERC) of Canada Discovery Grant and the Canada Research Chairs Program. 


\begin{thebibliography}{65}%
\makeatletter
\providecommand \@ifxundefined [1]{%
 \@ifx{#1\undefined}
}%
\providecommand \@ifnum [1]{%
 \ifnum #1\expandafter \@firstoftwo
 \else \expandafter \@secondoftwo
 \fi
}%
\providecommand \@ifx [1]{%
 \ifx #1\expandafter \@firstoftwo
 \else \expandafter \@secondoftwo
 \fi
}%
\providecommand \natexlab [1]{#1}%
\providecommand \enquote  [1]{``#1''}%
\providecommand \bibnamefont  [1]{#1}%
\providecommand \bibfnamefont [1]{#1}%
\providecommand \citenamefont [1]{#1}%
\providecommand \href@noop [0]{\@secondoftwo}%
\providecommand \href [0]{\begingroup \@sanitize@url \@href}%
\providecommand \@href[1]{\@@startlink{#1}\@@href}%
\providecommand \@@href[1]{\endgroup#1\@@endlink}%
\providecommand \@sanitize@url [0]{\catcode `\\12\catcode `\$12\catcode
  `\&12\catcode `\#12\catcode `\^12\catcode `\_12\catcode `\%12\relax}%
\providecommand \@@startlink[1]{}%
\providecommand \@@endlink[0]{}%
\providecommand \url  [0]{\begingroup\@sanitize@url \@url }%
\providecommand \@url [1]{\endgroup\@href {#1}{\urlprefix }}%
\providecommand \urlprefix  [0]{URL }%
\providecommand \Eprint [0]{\href }%
\providecommand \doibase [0]{http://dx.doi.org/}%
\providecommand \selectlanguage [0]{\@gobble}%
\providecommand \bibinfo  [0]{\@secondoftwo}%
\providecommand \bibfield  [0]{\@secondoftwo}%
\providecommand \translation [1]{[#1]}%
\providecommand \BibitemOpen [0]{}%
\providecommand \bibitemStop [0]{}%
\providecommand \bibitemNoStop [0]{.\EOS\space}%
\providecommand \EOS [0]{\spacefactor3000\relax}%
\providecommand \BibitemShut  [1]{\csname bibitem#1\endcsname}%
\let\auto@bib@innerbib\@empty
\bibitem [{\citenamefont {Scovil}\ and\ \citenamefont
  {Schulz-DuBois}(1959)}]{Scovil.59.PRL}%
  \BibitemOpen
  \bibfield  {author} {\bibinfo {author} {\bibfnamefont {H.~E.~D.}\
  \bibnamefont {Scovil}}\ and\ \bibinfo {author} {\bibfnamefont {E.~O.}\
  \bibnamefont {Schulz-DuBois}},\ }\bibfield  {title} {\enquote {\bibinfo
  {title} {Three-level masers as heat engines},}\ }\href {\doibase
  10.1103/PhysRevLett.2.262} {\bibfield  {journal} {\bibinfo  {journal} {Phys.
  Rev. Lett.}\ }\textbf {\bibinfo {volume} {2}},\ \bibinfo {pages} {262}
  (\bibinfo {year} {1959})}\BibitemShut {NoStop}%
\bibitem [{\citenamefont {Geva}\ and\ \citenamefont
  {Kosloff}(1992)}]{Geva.92.JCP}%
  \BibitemOpen
  \bibfield  {author} {\bibinfo {author} {\bibfnamefont {E.}~\bibnamefont
  {Geva}}\ and\ \bibinfo {author} {\bibfnamefont {R.}~\bibnamefont {Kosloff}},\
  }\bibfield  {title} {\enquote {\bibinfo {title} {On the classical limit of
  quantum thermodynamics in finite time},}\ }\href {\doibase 10.1063/1.463909}
  {\bibfield  {journal} {\bibinfo  {journal} {J. Chem. Phys.}\ }\textbf
  {\bibinfo {volume} {97}},\ \bibinfo {pages} {4398} (\bibinfo {year}
  {1992})}\BibitemShut {NoStop}%
\bibitem [{\citenamefont {Gemmer}\ \emph {et~al.}(2009)\citenamefont {Gemmer},
  \citenamefont {Michel},\ and\ \citenamefont {Mahler}}]{Gemmer.09.NULL}%
  \BibitemOpen
  \bibfield  {author} {\bibinfo {author} {\bibfnamefont {J.}~\bibnamefont
  {Gemmer}}, \bibinfo {author} {\bibfnamefont {M.}~\bibnamefont {Michel}}, \
  and\ \bibinfo {author} {\bibfnamefont {G.}~\bibnamefont {Mahler}},\
  }\href@noop {} {\emph {\bibinfo {title} {Quantum Thermodynamics}}}\ (\bibinfo
   {publisher} {Springer-Verlag, Berlin},\ \bibinfo {year} {2009})\BibitemShut
  {NoStop}%
\bibitem [{\citenamefont {Seifert}(2012)}]{Seifert.12.RPP}%
  \BibitemOpen
  \bibfield  {author} {\bibinfo {author} {\bibfnamefont {U.}~\bibnamefont
  {Seifert}},\ }\bibfield  {title} {\enquote {\bibinfo {title} {Stochastic
  thermodynamics, fluctuation theorems and molecular machines},}\ }\href
  {http://stacks.iop.org/0034-4885/75/i=12/a=126001} {\bibfield  {journal}
  {\bibinfo  {journal} {Rep. Prog. Phys.}\ }\textbf {\bibinfo {volume} {75}},\
  \bibinfo {pages} {126001} (\bibinfo {year} {2012})}\BibitemShut {NoStop}%
\bibitem [{\citenamefont {Pekola}(2015)}]{Pekola.15.NP}%
  \BibitemOpen
  \bibfield  {author} {\bibinfo {author} {\bibfnamefont {J.~P.}\ \bibnamefont
  {Pekola}},\ }\bibfield  {title} {\enquote {\bibinfo {title} {Towards quantum
  thermodynamics in electronic circuits},}\ }\href
  {https://doi.org/10.1038/nphys3169} {\bibfield  {journal} {\bibinfo
  {journal} {Nat. Phys.}\ }\textbf {\bibinfo {volume} {11}},\ \bibinfo {pages}
  {118} (\bibinfo {year} {2015})}\BibitemShut {NoStop}%
\bibitem [{\citenamefont {Kosloff}(2013)}]{Kosloff.15.E}%
  \BibitemOpen
  \bibfield  {author} {\bibinfo {author} {\bibfnamefont {R.}~\bibnamefont
  {Kosloff}},\ }\bibfield  {title} {\enquote {\bibinfo {title} {Quantum
  thermodynamics: A dynamical viewpoint},}\ }\href {\doibase 10.3390/e15062100}
  {\bibfield  {journal} {\bibinfo  {journal} {Entropy}\ }\textbf {\bibinfo
  {volume} {15}},\ \bibinfo {pages} {2100} (\bibinfo {year}
  {2013})}\BibitemShut {NoStop}%
\bibitem [{\citenamefont {Goold}\ \emph {et~al.}(2016)\citenamefont {Goold},
  \citenamefont {Huber}, \citenamefont {Riera}, \citenamefont {del Rio},\ and\
  \citenamefont {Skrzypczyk}}]{Goold.16.JPA}%
  \BibitemOpen
  \bibfield  {author} {\bibinfo {author} {\bibfnamefont {J.}~\bibnamefont
  {Goold}}, \bibinfo {author} {\bibfnamefont {M.}~\bibnamefont {Huber}},
  \bibinfo {author} {\bibfnamefont {A.}~\bibnamefont {Riera}}, \bibinfo
  {author} {\bibfnamefont {L.}~\bibnamefont {del Rio}}, \ and\ \bibinfo
  {author} {\bibfnamefont {P.}~\bibnamefont {Skrzypczyk}},\ }\bibfield  {title}
  {\enquote {\bibinfo {title} {The role of quantum information in
  thermodynamics?a topical review},}\ }\href
  {http://stacks.iop.org/1751-8121/49/i=14/a=143001} {\bibfield  {journal}
  {\bibinfo  {journal} {J. Phys. A: Math. Theor.}\ }\textbf {\bibinfo {volume}
  {49}},\ \bibinfo {pages} {143001} (\bibinfo {year} {2016})}\BibitemShut
  {NoStop}%
\bibitem [{\citenamefont {Vinjanampathy}\ and\ \citenamefont
  {Anders}(2016)}]{Anders.16.CP}%
  \BibitemOpen
  \bibfield  {author} {\bibinfo {author} {\bibfnamefont {S.}~\bibnamefont
  {Vinjanampathy}}\ and\ \bibinfo {author} {\bibfnamefont {J.}~\bibnamefont
  {Anders}},\ }\bibfield  {title} {\enquote {\bibinfo {title} {Quantum
  thermodynamics},}\ }\href {\doibase 10.1080/00107514.2016.1201896} {\bibfield
   {journal} {\bibinfo  {journal} {Contemp. Phys.}\ }\textbf {\bibinfo {volume}
  {57}},\ \bibinfo {pages} {545} (\bibinfo {year} {2016})}\BibitemShut
  {NoStop}%
\bibitem [{\citenamefont {Benenti}\ \emph {et~al.}(2017)\citenamefont
  {Benenti}, \citenamefont {Casati}, \citenamefont {Saito},\ and\ \citenamefont
  {Whitney}}]{Benenti.17.PR}%
  \BibitemOpen
  \bibfield  {author} {\bibinfo {author} {\bibfnamefont {G.}~\bibnamefont
  {Benenti}}, \bibinfo {author} {\bibfnamefont {G.}~\bibnamefont {Casati}},
  \bibinfo {author} {\bibfnamefont {K.}~\bibnamefont {Saito}}, \ and\ \bibinfo
  {author} {\bibfnamefont {R.~S.}\ \bibnamefont {Whitney}},\ }\bibfield
  {title} {\enquote {\bibinfo {title} {Fundamental aspects of steady-state
  conversion of heat to work at the nanoscale},}\ }\href {\doibase
  https://doi.org/10.1016/j.physrep.2017.05.008} {\bibfield  {journal}
  {\bibinfo  {journal} {Phys. Rep.}\ }\textbf {\bibinfo {volume} {694}},\
  \bibinfo {pages} {1} (\bibinfo {year} {2017})}\BibitemShut {NoStop}%
\bibitem [{\citenamefont {Binder}\ \emph {et~al.}(2018)\citenamefont {Binder},
  \citenamefont {Correa}, \citenamefont {Gogolin}, \citenamefont {Anders},\
  and\ \citenamefont {Adesso}}]{Binder.18.NULL}%
  \BibitemOpen
  \bibfield  {author} {\bibinfo {author} {\bibfnamefont {F.}~\bibnamefont
  {Binder}}, \bibinfo {author} {\bibfnamefont {L.~A.}\ \bibnamefont {Correa}},
  \bibinfo {author} {\bibfnamefont {C.}~\bibnamefont {Gogolin}}, \bibinfo
  {author} {\bibfnamefont {J.}~\bibnamefont {Anders}}, \ and\ \bibinfo {author}
  {\bibfnamefont {G.}~\bibnamefont {Adesso}},\ }\href@noop {} {\emph {\bibinfo
  {title} {Thermodynamics in the Quantum Regime}}}\ (\bibinfo  {publisher}
  {Springer International},\ \bibinfo {year} {2018})\BibitemShut {NoStop}%
\bibitem [{\citenamefont {Seifert}(2019)}]{Seifert.19.ARCMP}%
  \BibitemOpen
  \bibfield  {author} {\bibinfo {author} {\bibfnamefont {U.}~\bibnamefont
  {Seifert}},\ }\bibfield  {title} {\enquote {\bibinfo {title} {From stochastic
  thermodynamics to thermodynamic inference},}\ }\href {\doibase
  10.1146/annurev-conmatphys-031218-013554} {\bibfield  {journal} {\bibinfo
  {journal} {Annu. Rev. Condens. Matter Phys.}\ }\textbf {\bibinfo {volume}
  {10}},\ \bibinfo {pages} {171} (\bibinfo {year} {2019})}\BibitemShut
  {NoStop}%
\bibitem [{\citenamefont {Brandner}\ \emph {et~al.}(2015)\citenamefont
  {Brandner}, \citenamefont {Saito},\ and\ \citenamefont
  {Seifert}}]{Brandner.15.PRX}%
  \BibitemOpen
  \bibfield  {author} {\bibinfo {author} {\bibfnamefont {K.}~\bibnamefont
  {Brandner}}, \bibinfo {author} {\bibfnamefont {K.}~\bibnamefont {Saito}}, \
  and\ \bibinfo {author} {\bibfnamefont {U.}~\bibnamefont {Seifert}},\
  }\bibfield  {title} {\enquote {\bibinfo {title} {Thermodynamics of micro- and
  nano-systems driven by periodic temperature variations},}\ }\href {\doibase
  10.1103/PhysRevX.5.031019} {\bibfield  {journal} {\bibinfo  {journal} {Phys.
  Rev. X}\ }\textbf {\bibinfo {volume} {5}},\ \bibinfo {pages} {031019}
  (\bibinfo {year} {2015})}\BibitemShut {NoStop}%
\bibitem [{\citenamefont {Brandner}\ and\ \citenamefont
  {Seifert}(2016)}]{Brandner.16.PRE}%
  \BibitemOpen
  \bibfield  {author} {\bibinfo {author} {\bibfnamefont {K.}~\bibnamefont
  {Brandner}}\ and\ \bibinfo {author} {\bibfnamefont {U.}~\bibnamefont
  {Seifert}},\ }\bibfield  {title} {\enquote {\bibinfo {title} {Periodic
  thermodynamics of open quantum systems},}\ }\href {\doibase
  10.1103/PhysRevE.93.062134} {\bibfield  {journal} {\bibinfo  {journal} {Phys.
  Rev. E}\ }\textbf {\bibinfo {volume} {93}},\ \bibinfo {pages} {062134}
  (\bibinfo {year} {2016})}\BibitemShut {NoStop}%
\bibitem [{\citenamefont {Proesmans}\ \emph {et~al.}(2016)\citenamefont
  {Proesmans}, \citenamefont {Dreher}, \citenamefont {Gavrilov}, \citenamefont
  {Bechhoefer},\ and\ \citenamefont {Van~den Broeck}}]{Proesmans.16.PRX}%
  \BibitemOpen
  \bibfield  {author} {\bibinfo {author} {\bibfnamefont {K.}~\bibnamefont
  {Proesmans}}, \bibinfo {author} {\bibfnamefont {Y.}~\bibnamefont {Dreher}},
  \bibinfo {author} {\bibfnamefont {M.}~\bibnamefont {Gavrilov}}, \bibinfo
  {author} {\bibfnamefont {J.}~\bibnamefont {Bechhoefer}}, \ and\ \bibinfo
  {author} {\bibfnamefont {C.}~\bibnamefont {Van~den Broeck}},\ }\bibfield
  {title} {\enquote {\bibinfo {title} {Brownian duet: A novel tale of
  thermodynamic efficiency},}\ }\href {\doibase 10.1103/PhysRevX.6.041010}
  {\bibfield  {journal} {\bibinfo  {journal} {Phys. Rev. X}\ }\textbf {\bibinfo
  {volume} {6}},\ \bibinfo {pages} {041010} (\bibinfo {year}
  {2016})}\BibitemShut {NoStop}%
\bibitem [{\citenamefont {Newman}\ \emph {et~al.}(2017)\citenamefont {Newman},
  \citenamefont {Mintert},\ and\ \citenamefont {Nazir}}]{Newman.17.PRE}%
  \BibitemOpen
  \bibfield  {author} {\bibinfo {author} {\bibfnamefont {D.}~\bibnamefont
  {Newman}}, \bibinfo {author} {\bibfnamefont {F.}~\bibnamefont {Mintert}}, \
  and\ \bibinfo {author} {\bibfnamefont {A.}~\bibnamefont {Nazir}},\ }\bibfield
   {title} {\enquote {\bibinfo {title} {Performance of a quantum heat engine at
  strong reservoir coupling},}\ }\href {\doibase 10.1103/PhysRevE.95.032139}
  {\bibfield  {journal} {\bibinfo  {journal} {Phys. Rev. E}\ }\textbf {\bibinfo
  {volume} {95}},\ \bibinfo {pages} {032139} (\bibinfo {year}
  {2017})}\BibitemShut {NoStop}%
\bibitem [{\citenamefont {Alecce}\ \emph {et~al.}(2015)\citenamefont {Alecce},
  \citenamefont {Galve}, \citenamefont {Gullo}, \citenamefont {Dell'Anna},
  \citenamefont {Plastina},\ and\ \citenamefont {Zambrini}}]{Alecce.15.NJP}%
  \BibitemOpen
  \bibfield  {author} {\bibinfo {author} {\bibfnamefont {A.}~\bibnamefont
  {Alecce}}, \bibinfo {author} {\bibfnamefont {F.}~\bibnamefont {Galve}},
  \bibinfo {author} {\bibfnamefont {N.~Lo}\ \bibnamefont {Gullo}}, \bibinfo
  {author} {\bibfnamefont {L.}~\bibnamefont {Dell'Anna}}, \bibinfo {author}
  {\bibfnamefont {F.}~\bibnamefont {Plastina}}, \ and\ \bibinfo {author}
  {\bibfnamefont {R.}~\bibnamefont {Zambrini}},\ }\bibfield  {title} {\enquote
  {\bibinfo {title} {Quantum otto cycle with inner friction: finite-time and
  disorder effects},}\ }\href {\doibase 10.1088/1367-2630/17/7/075007}
  {\bibfield  {journal} {\bibinfo  {journal} {New J. Phys.}\ }\textbf {\bibinfo
  {volume} {17}},\ \bibinfo {pages} {075007} (\bibinfo {year}
  {2015})}\BibitemShut {NoStop}%
\bibitem [{\citenamefont {Perarnau-Llobet}\ \emph {et~al.}(2018)\citenamefont
  {Perarnau-Llobet}, \citenamefont {Wilming}, \citenamefont {Riera},
  \citenamefont {Gallego},\ and\ \citenamefont
  {Eisert}}]{Perarnau-Llobet.18.PRL}%
  \BibitemOpen
  \bibfield  {author} {\bibinfo {author} {\bibfnamefont {M.}~\bibnamefont
  {Perarnau-Llobet}}, \bibinfo {author} {\bibfnamefont {H.}~\bibnamefont
  {Wilming}}, \bibinfo {author} {\bibfnamefont {A.}~\bibnamefont {Riera}},
  \bibinfo {author} {\bibfnamefont {R.}~\bibnamefont {Gallego}}, \ and\
  \bibinfo {author} {\bibfnamefont {J.}~\bibnamefont {Eisert}},\ }\bibfield
  {title} {\enquote {\bibinfo {title} {Strong coupling corrections in quantum
  thermodynamics},}\ }\href {\doibase 10.1103/PhysRevLett.120.120602}
  {\bibfield  {journal} {\bibinfo  {journal} {Phys. Rev. Lett.}\ }\textbf
  {\bibinfo {volume} {120}},\ \bibinfo {pages} {120602} (\bibinfo {year}
  {2018})}\BibitemShut {NoStop}%
\bibitem [{\citenamefont {Mohammady}\ and\ \citenamefont
  {Romito}(2019)}]{Mohammady.19.PRE}%
  \BibitemOpen
  \bibfield  {author} {\bibinfo {author} {\bibfnamefont {M.~H.}\ \bibnamefont
  {Mohammady}}\ and\ \bibinfo {author} {\bibfnamefont {A.}~\bibnamefont
  {Romito}},\ }\bibfield  {title} {\enquote {\bibinfo {title} {Efficiency of a
  cyclic quantum heat engine with finite-size baths},}\ }\href {\doibase
  10.1103/PhysRevE.100.012122} {\bibfield  {journal} {\bibinfo  {journal}
  {Phys. Rev. E}\ }\textbf {\bibinfo {volume} {100}},\ \bibinfo {pages}
  {012122} (\bibinfo {year} {2019})}\BibitemShut {NoStop}%
\bibitem [{\citenamefont {Kloc}\ \emph {et~al.}(2019)\citenamefont {Kloc},
  \citenamefont {Cejnar},\ and\ \citenamefont {Schaller}}]{Kloc.19.PRE}%
  \BibitemOpen
  \bibfield  {author} {\bibinfo {author} {\bibfnamefont {M.}~\bibnamefont
  {Kloc}}, \bibinfo {author} {\bibfnamefont {P.}~\bibnamefont {Cejnar}}, \ and\
  \bibinfo {author} {\bibfnamefont {G.}~\bibnamefont {Schaller}},\ }\bibfield
  {title} {\enquote {\bibinfo {title} {Collective performance of a finite-time
  quantum otto cycle},}\ }\href {\doibase 10.1103/PhysRevE.100.042126}
  {\bibfield  {journal} {\bibinfo  {journal} {Phys. Rev. E}\ }\textbf {\bibinfo
  {volume} {100}},\ \bibinfo {pages} {042126} (\bibinfo {year}
  {2019})}\BibitemShut {NoStop}%
\bibitem [{\citenamefont {Pezzutto}\ \emph {et~al.}(2019)\citenamefont
  {Pezzutto}, \citenamefont {Paternostro},\ and\ \citenamefont
  {Omar}}]{Pezzutto.19.QST}%
  \BibitemOpen
  \bibfield  {author} {\bibinfo {author} {\bibfnamefont {M.}~\bibnamefont
  {Pezzutto}}, \bibinfo {author} {\bibfnamefont {M.}~\bibnamefont
  {Paternostro}}, \ and\ \bibinfo {author} {\bibfnamefont {Y.}~\bibnamefont
  {Omar}},\ }\bibfield  {title} {\enquote {\bibinfo {title} {An
  out-of-equilibrium non-markovian quantum heat engine},}\ }\href {\doibase
  10.1088/2058-9565/aaf5b4} {\bibfield  {journal} {\bibinfo  {journal} {Quantum
  Sci. Technol.}\ }\textbf {\bibinfo {volume} {4}},\ \bibinfo {pages} {025002}
  (\bibinfo {year} {2019})}\BibitemShut {NoStop}%
\bibitem [{\citenamefont {Park}\ \emph {et~al.}(2019)\citenamefont {Park},
  \citenamefont {Lee}, \citenamefont {Chun},\ and\ \citenamefont
  {Noh}}]{ParkJ.19.PRE}%
  \BibitemOpen
  \bibfield  {author} {\bibinfo {author} {\bibfnamefont {J-M.}\ \bibnamefont
  {Park}}, \bibinfo {author} {\bibfnamefont {S.}~\bibnamefont {Lee}}, \bibinfo
  {author} {\bibfnamefont {H-M.}\ \bibnamefont {Chun}}, \ and\ \bibinfo
  {author} {\bibfnamefont {J.~D.}\ \bibnamefont {Noh}},\ }\bibfield  {title}
  {\enquote {\bibinfo {title} {Quantum mechanical bound for efficiency of
  quantum otto heat engine},}\ }\href {\doibase 10.1103/PhysRevE.100.012148}
  {\bibfield  {journal} {\bibinfo  {journal} {Phys. Rev. E}\ }\textbf {\bibinfo
  {volume} {100}},\ \bibinfo {pages} {012148} (\bibinfo {year}
  {2019})}\BibitemShut {NoStop}%
\bibitem [{\citenamefont {Solfanelli}\ \emph {et~al.}(2020)\citenamefont
  {Solfanelli}, \citenamefont {Falsetti},\ and\ \citenamefont
  {Campisi}}]{Solfanelli.20.PRB}%
  \BibitemOpen
  \bibfield  {author} {\bibinfo {author} {\bibfnamefont {A.}~\bibnamefont
  {Solfanelli}}, \bibinfo {author} {\bibfnamefont {M.}~\bibnamefont
  {Falsetti}}, \ and\ \bibinfo {author} {\bibfnamefont {M.}~\bibnamefont
  {Campisi}},\ }\bibfield  {title} {\enquote {\bibinfo {title} {Nonadiabatic
  single-qubit quantum otto engine},}\ }\href {\doibase
  10.1103/PhysRevB.101.054513} {\bibfield  {journal} {\bibinfo  {journal}
  {Phys. Rev. B}\ }\textbf {\bibinfo {volume} {101}},\ \bibinfo {pages}
  {054513} (\bibinfo {year} {2020})}\BibitemShut {NoStop}%
\bibitem [{\citenamefont {Çakmak}\ and\ \citenamefont
  {Altintas}(2020)}]{Altintas.20.QIP}%
  \BibitemOpen
  \bibfield  {author} {\bibinfo {author} {\bibfnamefont {S.}~\bibnamefont
  {Çakmak}}\ and\ \bibinfo {author} {\bibfnamefont {F.}~\bibnamefont
  {Altintas}},\ }\bibfield  {title} {\enquote {\bibinfo {title} {Quantum carnot
  cycle with inner friction},}\ }\href
  {https://doi.org/10.1007/s11128-020-02746-x} {\bibfield  {journal} {\bibinfo
  {journal} {Quantum Inf. Processing}\ }\textbf {\bibinfo {volume} {19}},\
  \bibinfo {pages} {248} (\bibinfo {year} {2020})}\BibitemShut {NoStop}%
\bibitem [{\citenamefont {Wiedmann}\ \emph {et~al.}(2020)\citenamefont
  {Wiedmann}, \citenamefont {Stockburger},\ and\ \citenamefont
  {Ankerhold}}]{Wiedmann.20.NJP}%
  \BibitemOpen
  \bibfield  {author} {\bibinfo {author} {\bibfnamefont {M.}~\bibnamefont
  {Wiedmann}}, \bibinfo {author} {\bibfnamefont {J.~T.}\ \bibnamefont
  {Stockburger}}, \ and\ \bibinfo {author} {\bibfnamefont {J.}~\bibnamefont
  {Ankerhold}},\ }\bibfield  {title} {\enquote {\bibinfo {title} {Non-markovian
  dynamics of a quantum heat engine: out-of-equilibrium operation and thermal
  coupling control},}\ }\href {\doibase 10.1088/1367-2630/ab725a} {\bibfield
  {journal} {\bibinfo  {journal} {New J. Phys.}\ }\textbf {\bibinfo {volume}
  {22}},\ \bibinfo {pages} {033007} (\bibinfo {year} {2020})}\BibitemShut
  {NoStop}%
\bibitem [{\citenamefont {Shirai}\ \emph {et~al.}(2021)\citenamefont {Shirai},
  \citenamefont {Hashimoto}, \citenamefont {Tezuka}, \citenamefont {Uchiyama},\
  and\ \citenamefont {Hatano}}]{Shirai.21.PRR}%
  \BibitemOpen
  \bibfield  {author} {\bibinfo {author} {\bibfnamefont {Y.}~\bibnamefont
  {Shirai}}, \bibinfo {author} {\bibfnamefont {K.}~\bibnamefont {Hashimoto}},
  \bibinfo {author} {\bibfnamefont {R.}~\bibnamefont {Tezuka}}, \bibinfo
  {author} {\bibfnamefont {C.}~\bibnamefont {Uchiyama}}, \ and\ \bibinfo
  {author} {\bibfnamefont {N.}~\bibnamefont {Hatano}},\ }\bibfield  {title}
  {\enquote {\bibinfo {title} {Non-markovian effect on quantum otto engine:
  Role of system-reservoir interaction},}\ }\href {\doibase
  10.1103/PhysRevResearch.3.023078} {\bibfield  {journal} {\bibinfo  {journal}
  {Phys. Rev. Research}\ }\textbf {\bibinfo {volume} {3}},\ \bibinfo {pages}
  {023078} (\bibinfo {year} {2021})}\BibitemShut {NoStop}%
\bibitem [{\citenamefont {Strasberg}\ \emph {et~al.}(2021)\citenamefont
  {Strasberg}, \citenamefont {W\"achtler},\ and\ \citenamefont
  {Schaller}}]{Strasberg.21.PRL}%
  \BibitemOpen
  \bibfield  {author} {\bibinfo {author} {\bibfnamefont {P.}~\bibnamefont
  {Strasberg}}, \bibinfo {author} {\bibfnamefont {C.~W.}\ \bibnamefont
  {W\"achtler}}, \ and\ \bibinfo {author} {\bibfnamefont {G.}~\bibnamefont
  {Schaller}},\ }\bibfield  {title} {\enquote {\bibinfo {title} {Autonomous
  implementation of thermodynamic cycles at the nanoscale},}\ }\href {\doibase
  10.1103/PhysRevLett.126.180605} {\bibfield  {journal} {\bibinfo  {journal}
  {Phys. Rev. Lett.}\ }\textbf {\bibinfo {volume} {126}},\ \bibinfo {pages}
  {180605} (\bibinfo {year} {2021})}\BibitemShut {NoStop}%
\bibitem [{\citenamefont {Abah}\ \emph {et~al.}(2012)\citenamefont {Abah},
  \citenamefont {Ro\ss{}nagel}, \citenamefont {Jacob}, \citenamefont {Deffner},
  \citenamefont {Schmidt-Kaler}, \citenamefont {Singer},\ and\ \citenamefont
  {Lutz}}]{Abah.12.PRL}%
  \BibitemOpen
  \bibfield  {author} {\bibinfo {author} {\bibfnamefont {O.}~\bibnamefont
  {Abah}}, \bibinfo {author} {\bibfnamefont {J.}~\bibnamefont {Ro\ss{}nagel}},
  \bibinfo {author} {\bibfnamefont {G.}~\bibnamefont {Jacob}}, \bibinfo
  {author} {\bibfnamefont {S.}~\bibnamefont {Deffner}}, \bibinfo {author}
  {\bibfnamefont {F.}~\bibnamefont {Schmidt-Kaler}}, \bibinfo {author}
  {\bibfnamefont {K.}~\bibnamefont {Singer}}, \ and\ \bibinfo {author}
  {\bibfnamefont {E.}~\bibnamefont {Lutz}},\ }\bibfield  {title} {\enquote
  {\bibinfo {title} {Single-ion heat engine at maximum power},}\ }\href
  {\doibase 10.1103/PhysRevLett.109.203006} {\bibfield  {journal} {\bibinfo
  {journal} {Phys. Rev. Lett.}\ }\textbf {\bibinfo {volume} {109}},\ \bibinfo
  {pages} {203006} (\bibinfo {year} {2012})}\BibitemShut {NoStop}%
\bibitem [{\citenamefont {Ro\ss{}nagel}\ \emph {et~al.}(2014)\citenamefont
  {Ro\ss{}nagel}, \citenamefont {Abah}, \citenamefont {Schmidt-Kaler},
  \citenamefont {Singer},\ and\ \citenamefont {Lutz}}]{Rossnagel.14.PRL}%
  \BibitemOpen
  \bibfield  {author} {\bibinfo {author} {\bibfnamefont {J.}~\bibnamefont
  {Ro\ss{}nagel}}, \bibinfo {author} {\bibfnamefont {O.}~\bibnamefont {Abah}},
  \bibinfo {author} {\bibfnamefont {F.}~\bibnamefont {Schmidt-Kaler}}, \bibinfo
  {author} {\bibfnamefont {K.}~\bibnamefont {Singer}}, \ and\ \bibinfo {author}
  {\bibfnamefont {E.}~\bibnamefont {Lutz}},\ }\bibfield  {title} {\enquote
  {\bibinfo {title} {Nanoscale heat engine beyond the carnot limit},}\ }\href
  {\doibase 10.1103/PhysRevLett.112.030602} {\bibfield  {journal} {\bibinfo
  {journal} {Phys. Rev. Lett.}\ }\textbf {\bibinfo {volume} {112}},\ \bibinfo
  {pages} {030602} (\bibinfo {year} {2014})}\BibitemShut {NoStop}%
\bibitem [{\citenamefont {Dechant}\ \emph {et~al.}(2015)\citenamefont
  {Dechant}, \citenamefont {Kiesel},\ and\ \citenamefont
  {Lutz}}]{Dechant.15.PRL}%
  \BibitemOpen
  \bibfield  {author} {\bibinfo {author} {\bibfnamefont {A.}~\bibnamefont
  {Dechant}}, \bibinfo {author} {\bibfnamefont {N.}~\bibnamefont {Kiesel}}, \
  and\ \bibinfo {author} {\bibfnamefont {E.}~\bibnamefont {Lutz}},\ }\bibfield
  {title} {\enquote {\bibinfo {title} {All-optical nanomechanical heat
  engine},}\ }\href {\doibase 10.1103/PhysRevLett.114.183602} {\bibfield
  {journal} {\bibinfo  {journal} {Phys. Rev. Lett.}\ }\textbf {\bibinfo
  {volume} {114}},\ \bibinfo {pages} {183602} (\bibinfo {year}
  {2015})}\BibitemShut {NoStop}%
\bibitem [{\citenamefont {Krishnamurthy}\ \emph {et~al.}(2016)\citenamefont
  {Krishnamurthy}, \citenamefont {Ghosh}, \citenamefont {Chatterji},
  \citenamefont {Ganapathy},\ and\ \citenamefont {Sood}}]{Sood.16.NP}%
  \BibitemOpen
  \bibfield  {author} {\bibinfo {author} {\bibfnamefont {S.}~\bibnamefont
  {Krishnamurthy}}, \bibinfo {author} {\bibfnamefont {S.}~\bibnamefont
  {Ghosh}}, \bibinfo {author} {\bibfnamefont {D.}~\bibnamefont {Chatterji}},
  \bibinfo {author} {\bibfnamefont {R.}~\bibnamefont {Ganapathy}}, \ and\
  \bibinfo {author} {\bibfnamefont {A.~K.}\ \bibnamefont {Sood}},\ }\bibfield
  {title} {\enquote {\bibinfo {title} {A micrometre-sized heat engine operating
  between bacterial reservoirs},}\ }\href {https://doi.org/10.1038/nphys3870}
  {\bibfield  {journal} {\bibinfo  {journal} {Nat. Phys.}\ }\textbf {\bibinfo
  {volume} {12}},\ \bibinfo {pages} {1134} (\bibinfo {year}
  {2016})}\BibitemShut {NoStop}%
\bibitem [{\citenamefont {Roßnagel}\ \emph {et~al.}(2016)\citenamefont
  {Roßnagel}, \citenamefont {Dawkins}, \citenamefont {Tolazzi}, \citenamefont
  {Abah}, \citenamefont {Lutz}, \citenamefont {Schmidt-Kaler},\ and\
  \citenamefont {Singer}}]{Robnagel.16.S}%
  \BibitemOpen
  \bibfield  {author} {\bibinfo {author} {\bibfnamefont {J.}~\bibnamefont
  {Roßnagel}}, \bibinfo {author} {\bibfnamefont {Samuel~T.}\ \bibnamefont
  {Dawkins}}, \bibinfo {author} {\bibfnamefont {K.~N.}\ \bibnamefont
  {Tolazzi}}, \bibinfo {author} {\bibfnamefont {O.}~\bibnamefont {Abah}},
  \bibinfo {author} {\bibfnamefont {E.}~\bibnamefont {Lutz}}, \bibinfo {author}
  {\bibfnamefont {F.}~\bibnamefont {Schmidt-Kaler}}, \ and\ \bibinfo {author}
  {\bibfnamefont {K.}~\bibnamefont {Singer}},\ }\bibfield  {title} {\enquote
  {\bibinfo {title} {A single-atom heat engine},}\ }\href
  {http://science.sciencemag.org/content/352/6283/325.abstract} {\bibfield
  {journal} {\bibinfo  {journal} {Science}\ }\textbf {\bibinfo {volume}
  {352}},\ \bibinfo {pages} {325} (\bibinfo {year} {2016})}\BibitemShut
  {NoStop}%
\bibitem [{\citenamefont {Peterson}\ \emph {et~al.}(2019)\citenamefont
  {Peterson}, \citenamefont {Batalh\~ao}, \citenamefont {Herrera},
  \citenamefont {Souza}, \citenamefont {Sarthour}, \citenamefont {Oliveira},\
  and\ \citenamefont {Serra}}]{Peterson.19.PRL}%
  \BibitemOpen
  \bibfield  {author} {\bibinfo {author} {\bibfnamefont {J.~P.~S.}\
  \bibnamefont {Peterson}}, \bibinfo {author} {\bibfnamefont {T.~B.}\
  \bibnamefont {Batalh\~ao}}, \bibinfo {author} {\bibfnamefont
  {M.}~\bibnamefont {Herrera}}, \bibinfo {author} {\bibfnamefont {A.~M.}\
  \bibnamefont {Souza}}, \bibinfo {author} {\bibfnamefont {R.~S.}\ \bibnamefont
  {Sarthour}}, \bibinfo {author} {\bibfnamefont {I.~S.}\ \bibnamefont
  {Oliveira}}, \ and\ \bibinfo {author} {\bibfnamefont {R.~M.}\ \bibnamefont
  {Serra}},\ }\bibfield  {title} {\enquote {\bibinfo {title} {Experimental
  characterization of a spin quantum heat engine},}\ }\href {\doibase
  10.1103/PhysRevLett.123.240601} {\bibfield  {journal} {\bibinfo  {journal}
  {Phys. Rev. Lett.}\ }\textbf {\bibinfo {volume} {123}},\ \bibinfo {pages}
  {240601} (\bibinfo {year} {2019})}\BibitemShut {NoStop}%
\bibitem [{\citenamefont {Klatzow}\ \emph {et~al.}(2019)\citenamefont
  {Klatzow}, \citenamefont {Becker}, \citenamefont {Ledingham}, \citenamefont
  {Weinzetl}, \citenamefont {Kaczmarek}, \citenamefont {Saunders},
  \citenamefont {Nunn}, \citenamefont {Walmsley}, \citenamefont {Uzdin},\ and\
  \citenamefont {Poem}}]{Klatzow.19.PRL}%
  \BibitemOpen
  \bibfield  {author} {\bibinfo {author} {\bibfnamefont {J.}~\bibnamefont
  {Klatzow}}, \bibinfo {author} {\bibfnamefont {J.~N.}\ \bibnamefont {Becker}},
  \bibinfo {author} {\bibfnamefont {P.~M.}\ \bibnamefont {Ledingham}}, \bibinfo
  {author} {\bibfnamefont {C.}~\bibnamefont {Weinzetl}}, \bibinfo {author}
  {\bibfnamefont {K.~T.}\ \bibnamefont {Kaczmarek}}, \bibinfo {author}
  {\bibfnamefont {D.~J.}\ \bibnamefont {Saunders}}, \bibinfo {author}
  {\bibfnamefont {J.}~\bibnamefont {Nunn}}, \bibinfo {author} {\bibfnamefont
  {I.~A.}\ \bibnamefont {Walmsley}}, \bibinfo {author} {\bibfnamefont
  {R.}~\bibnamefont {Uzdin}}, \ and\ \bibinfo {author} {\bibfnamefont
  {E.}~\bibnamefont {Poem}},\ }\bibfield  {title} {\enquote {\bibinfo {title}
  {Experimental demonstration of quantum effects in the operation of
  microscopic heat engines},}\ }\href {\doibase 10.1103/PhysRevLett.122.110601}
  {\bibfield  {journal} {\bibinfo  {journal} {Phys. Rev. Lett.}\ }\textbf
  {\bibinfo {volume} {122}},\ \bibinfo {pages} {110601} (\bibinfo {year}
  {2019})}\BibitemShut {NoStop}%
\bibitem [{\citenamefont {de~Assis}\ \emph {et~al.}(2019)\citenamefont
  {de~Assis}, \citenamefont {de~Mendon\ifmmode~\mbox{\c{c}}\else \c{c}\fi{}a},
  \citenamefont {Villas-Boas}, \citenamefont {de~Souza}, \citenamefont
  {Sarthour}, \citenamefont {Oliveira},\ and\ \citenamefont
  {de~Almeida}}]{Assis.19.PRL}%
  \BibitemOpen
  \bibfield  {author} {\bibinfo {author} {\bibfnamefont {R.~J.}\ \bibnamefont
  {de~Assis}}, \bibinfo {author} {\bibfnamefont {T.~M.}\ \bibnamefont
  {de~Mendon\ifmmode~\mbox{\c{c}}\else \c{c}\fi{}a}}, \bibinfo {author}
  {\bibfnamefont {C.~J.}\ \bibnamefont {Villas-Boas}}, \bibinfo {author}
  {\bibfnamefont {A.~M.}\ \bibnamefont {de~Souza}}, \bibinfo {author}
  {\bibfnamefont {R.~S.}\ \bibnamefont {Sarthour}}, \bibinfo {author}
  {\bibfnamefont {I.~S.}\ \bibnamefont {Oliveira}}, \ and\ \bibinfo {author}
  {\bibfnamefont {N.~G.}\ \bibnamefont {de~Almeida}},\ }\bibfield  {title}
  {\enquote {\bibinfo {title} {Efficiency of a quantum otto heat engine
  operating under a reservoir at effective negative temperatures},}\ }\href
  {\doibase 10.1103/PhysRevLett.122.240602} {\bibfield  {journal} {\bibinfo
  {journal} {Phys. Rev. Lett.}\ }\textbf {\bibinfo {volume} {122}},\ \bibinfo
  {pages} {240602} (\bibinfo {year} {2019})}\BibitemShut {NoStop}%
\bibitem [{\citenamefont {Pekola}\ and\ \citenamefont
  {Khaymovich}(2019)}]{Pekola.19.ARCMP}%
  \BibitemOpen
  \bibfield  {author} {\bibinfo {author} {\bibfnamefont {J.~P.}\ \bibnamefont
  {Pekola}}\ and\ \bibinfo {author} {\bibfnamefont {I.~M.}\ \bibnamefont
  {Khaymovich}},\ }\bibfield  {title} {\enquote {\bibinfo {title}
  {Thermodynamics in single-electron circuits and superconducting qubits},}\
  }\href {\doibase 10.1146/annurev-conmatphys-033117-054120} {\bibfield
  {journal} {\bibinfo  {journal} {Annu. Rev. Condens. Matter Phys.}\ }\textbf
  {\bibinfo {volume} {10}},\ \bibinfo {pages} {193} (\bibinfo {year}
  {2019})}\BibitemShut {NoStop}%
\bibitem [{\citenamefont {von Lindenfels}\ \emph {et~al.}(2019)\citenamefont
  {von Lindenfels}, \citenamefont {Gr\"ab}, \citenamefont {Schmiegelow},
  \citenamefont {Kaushal}, \citenamefont {Schulz}, \citenamefont {Mitchison},
  \citenamefont {Goold}, \citenamefont {Schmidt-Kaler},\ and\ \citenamefont
  {Poschinger}}]{Lindenfels.19.PRL}%
  \BibitemOpen
  \bibfield  {author} {\bibinfo {author} {\bibfnamefont {D.}~\bibnamefont {von
  Lindenfels}}, \bibinfo {author} {\bibfnamefont {O.}~\bibnamefont {Gr\"ab}},
  \bibinfo {author} {\bibfnamefont {C.~T.}\ \bibnamefont {Schmiegelow}},
  \bibinfo {author} {\bibfnamefont {V.}~\bibnamefont {Kaushal}}, \bibinfo
  {author} {\bibfnamefont {J.}~\bibnamefont {Schulz}}, \bibinfo {author}
  {\bibfnamefont {Mark~T.}\ \bibnamefont {Mitchison}}, \bibinfo {author}
  {\bibfnamefont {John}\ \bibnamefont {Goold}}, \bibinfo {author}
  {\bibfnamefont {F.}~\bibnamefont {Schmidt-Kaler}}, \ and\ \bibinfo {author}
  {\bibfnamefont {U.~G.}\ \bibnamefont {Poschinger}},\ }\bibfield  {title}
  {\enquote {\bibinfo {title} {Spin heat engine coupled to a
  harmonic-oscillator flywheel},}\ }\href {\doibase
  10.1103/PhysRevLett.123.080602} {\bibfield  {journal} {\bibinfo  {journal}
  {Phys. Rev. Lett.}\ }\textbf {\bibinfo {volume} {123}},\ \bibinfo {pages}
  {080602} (\bibinfo {year} {2019})}\BibitemShut {NoStop}%
\bibitem [{\citenamefont {Feldmann}\ and\ \citenamefont
  {Kosloff}(2004)}]{Feldmann.04.PRE}%
  \BibitemOpen
  \bibfield  {author} {\bibinfo {author} {\bibfnamefont {T.}~\bibnamefont
  {Feldmann}}\ and\ \bibinfo {author} {\bibfnamefont {R.}~\bibnamefont
  {Kosloff}},\ }\bibfield  {title} {\enquote {\bibinfo {title} {Characteristics
  of the limit cycle of a reciprocating quantum heat engine},}\ }\href
  {\doibase 10.1103/PhysRevE.70.046110} {\bibfield  {journal} {\bibinfo
  {journal} {Phys. Rev. E}\ }\textbf {\bibinfo {volume} {70}},\ \bibinfo
  {pages} {046110} (\bibinfo {year} {2004})}\BibitemShut {NoStop}%
\bibitem [{\citenamefont {Insinga}\ \emph {et~al.}(2016)\citenamefont
  {Insinga}, \citenamefont {Andresen},\ and\ \citenamefont
  {Salamon}}]{Insinga.16.PRE}%
  \BibitemOpen
  \bibfield  {author} {\bibinfo {author} {\bibfnamefont {A.}~\bibnamefont
  {Insinga}}, \bibinfo {author} {\bibfnamefont {B.}~\bibnamefont {Andresen}}, \
  and\ \bibinfo {author} {\bibfnamefont {P.}~\bibnamefont {Salamon}},\
  }\bibfield  {title} {\enquote {\bibinfo {title} {Thermodynamical analysis of
  a quantum heat engine based on harmonic oscillators},}\ }\href {\doibase
  10.1103/PhysRevE.94.012119} {\bibfield  {journal} {\bibinfo  {journal} {Phys.
  Rev. E}\ }\textbf {\bibinfo {volume} {94}},\ \bibinfo {pages} {012119}
  (\bibinfo {year} {2016})}\BibitemShut {NoStop}%
\bibitem [{\citenamefont {Insinga}\ \emph {et~al.}(2018)\citenamefont
  {Insinga}, \citenamefont {Andresen}, \citenamefont {Salamon},\ and\
  \citenamefont {Kosloff}}]{Insinga.18.PRE}%
  \BibitemOpen
  \bibfield  {author} {\bibinfo {author} {\bibfnamefont {A.}~\bibnamefont
  {Insinga}}, \bibinfo {author} {\bibfnamefont {B.}~\bibnamefont {Andresen}},
  \bibinfo {author} {\bibfnamefont {P.}~\bibnamefont {Salamon}}, \ and\
  \bibinfo {author} {\bibfnamefont {R.}~\bibnamefont {Kosloff}},\ }\bibfield
  {title} {\enquote {\bibinfo {title} {Quantum heat engines: Limit cycles and
  exceptional points},}\ }\href {\doibase 10.1103/PhysRevE.97.062153}
  {\bibfield  {journal} {\bibinfo  {journal} {Phys. Rev. E}\ }\textbf {\bibinfo
  {volume} {97}},\ \bibinfo {pages} {062153} (\bibinfo {year}
  {2018})}\BibitemShut {NoStop}%
\bibitem [{\citenamefont {Menczel}\ and\ \citenamefont
  {Brandner}(2019)}]{Menczel.19.JPA}%
  \BibitemOpen
  \bibfield  {author} {\bibinfo {author} {\bibfnamefont {P.}~\bibnamefont
  {Menczel}}\ and\ \bibinfo {author} {\bibfnamefont {K.}~\bibnamefont
  {Brandner}},\ }\bibfield  {title} {\enquote {\bibinfo {title} {Limit cycles
  in periodically driven open quantum systems},}\ }\href {\doibase
  10.1088/1751-8121/ab435a} {\bibfield  {journal} {\bibinfo  {journal} {J.
  Phys. A: Math. Theor.}\ }\textbf {\bibinfo {volume} {52}},\ \bibinfo {pages}
  {43LT01} (\bibinfo {year} {2019})}\BibitemShut {NoStop}%
\bibitem [{\citenamefont {Talkner}\ and\ \citenamefont
  {H\"anggi}(2020)}]{Talkner.20.RMP}%
  \BibitemOpen
  \bibfield  {author} {\bibinfo {author} {\bibfnamefont {P.}~\bibnamefont
  {Talkner}}\ and\ \bibinfo {author} {\bibfnamefont {P.}~\bibnamefont
  {H\"anggi}},\ }\bibfield  {title} {\enquote {\bibinfo {title} {Colloquium:
  Statistical mechanics and thermodynamics at strong coupling: Quantum and
  classical},}\ }\href {\doibase 10.1103/RevModPhys.92.041002} {\bibfield
  {journal} {\bibinfo  {journal} {Rev. Mod. Phys.}\ }\textbf {\bibinfo {volume}
  {92}},\ \bibinfo {pages} {041002} (\bibinfo {year} {2020})}\BibitemShut
  {NoStop}%
\bibitem [{\citenamefont {Seshadri}\ and\ \citenamefont
  {Galperin}(2021)}]{Seshadri.21.PRB}%
  \BibitemOpen
  \bibfield  {author} {\bibinfo {author} {\bibfnamefont {N.}~\bibnamefont
  {Seshadri}}\ and\ \bibinfo {author} {\bibfnamefont {M.}~\bibnamefont
  {Galperin}},\ }\bibfield  {title} {\enquote {\bibinfo {title} {Entropy and
  information flow in quantum systems strongly coupled to baths},}\ }\href
  {\doibase 10.1103/PhysRevB.103.085415} {\bibfield  {journal} {\bibinfo
  {journal} {Phys. Rev. B}\ }\textbf {\bibinfo {volume} {103}},\ \bibinfo
  {pages} {085415} (\bibinfo {year} {2021})}\BibitemShut {NoStop}%
\bibitem [{\citenamefont {Pancotti}\ \emph {et~al.}(2020)\citenamefont
  {Pancotti}, \citenamefont {Scandi}, \citenamefont {Mitchison},\ and\
  \citenamefont {Perarnau-Llobet}}]{Pancotti.20.PRX}%
  \BibitemOpen
  \bibfield  {author} {\bibinfo {author} {\bibfnamefont {N.}~\bibnamefont
  {Pancotti}}, \bibinfo {author} {\bibfnamefont {M.}~\bibnamefont {Scandi}},
  \bibinfo {author} {\bibfnamefont {M.~T.}\ \bibnamefont {Mitchison}}, \ and\
  \bibinfo {author} {\bibfnamefont {M.}~\bibnamefont {Perarnau-Llobet}},\
  }\bibfield  {title} {\enquote {\bibinfo {title} {Speed-ups to isothermality:
  Enhanced quantum thermal machines through control of the system-bath
  coupling},}\ }\href {\doibase 10.1103/PhysRevX.10.031015} {\bibfield
  {journal} {\bibinfo  {journal} {Phys. Rev. X}\ }\textbf {\bibinfo {volume}
  {10}},\ \bibinfo {pages} {031015} (\bibinfo {year} {2020})}\BibitemShut
  {NoStop}%
\bibitem [{\citenamefont {Bruch}\ \emph {et~al.}(2016)\citenamefont {Bruch},
  \citenamefont {Thomas}, \citenamefont {Viola~Kusminskiy}, \citenamefont {von
  Oppen},\ and\ \citenamefont {Nitzan}}]{Bruch.16.PRB}%
  \BibitemOpen
  \bibfield  {author} {\bibinfo {author} {\bibfnamefont {A.}~\bibnamefont
  {Bruch}}, \bibinfo {author} {\bibfnamefont {M.}~\bibnamefont {Thomas}},
  \bibinfo {author} {\bibfnamefont {S.}~\bibnamefont {Viola~Kusminskiy}},
  \bibinfo {author} {\bibfnamefont {F.}~\bibnamefont {von Oppen}}, \ and\
  \bibinfo {author} {\bibfnamefont {A.}~\bibnamefont {Nitzan}},\ }\bibfield
  {title} {\enquote {\bibinfo {title} {Quantum thermodynamics of the driven
  resonant level model},}\ }\href {\doibase 10.1103/PhysRevB.93.115318}
  {\bibfield  {journal} {\bibinfo  {journal} {Phys. Rev. B}\ }\textbf {\bibinfo
  {volume} {93}},\ \bibinfo {pages} {115318} (\bibinfo {year}
  {2016})}\BibitemShut {NoStop}%
\bibitem [{\citenamefont {Oz}\ \emph {et~al.}(2020)\citenamefont {Oz},
  \citenamefont {Hod},\ and\ \citenamefont {Nitzan}}]{Oz.20.JCTC}%
  \BibitemOpen
  \bibfield  {author} {\bibinfo {author} {\bibfnamefont {A.}~\bibnamefont
  {Oz}}, \bibinfo {author} {\bibfnamefont {O.}~\bibnamefont {Hod}}, \ and\
  \bibinfo {author} {\bibfnamefont {A.}~\bibnamefont {Nitzan}},\ }\bibfield
  {title} {\enquote {\bibinfo {title} {Numerical approach to nonequilibrium
  quantum thermodynamics: Nonperturbative treatment of the driven resonant
  level model based on the driven liouville von-neumann formalism},}\ }\href
  {\doibase 10.1021/acs.jctc.9b00999} {\bibfield  {journal} {\bibinfo
  {journal} {J. Chem. Theory Comput.}\ }\textbf {\bibinfo {volume} {16}},\
  \bibinfo {pages} {1232} (\bibinfo {year} {2020})}\BibitemShut {NoStop}%
\bibitem [{\citenamefont {Cangemi}\ \emph {et~al.}(2021)\citenamefont
  {Cangemi}, \citenamefont {Carrega}, \citenamefont {De~Candia}, \citenamefont
  {Cataudella}, \citenamefont {De~Filippis}, \citenamefont {Sassetti},\ and\
  \citenamefont {Benenti}}]{Cangemi.21.PRR}%
  \BibitemOpen
  \bibfield  {author} {\bibinfo {author} {\bibfnamefont {L.~M.}\ \bibnamefont
  {Cangemi}}, \bibinfo {author} {\bibfnamefont {M.}~\bibnamefont {Carrega}},
  \bibinfo {author} {\bibfnamefont {A.}~\bibnamefont {De~Candia}}, \bibinfo
  {author} {\bibfnamefont {V.}~\bibnamefont {Cataudella}}, \bibinfo {author}
  {\bibfnamefont {G.}~\bibnamefont {De~Filippis}}, \bibinfo {author}
  {\bibfnamefont {M.}~\bibnamefont {Sassetti}}, \ and\ \bibinfo {author}
  {\bibfnamefont {G.}~\bibnamefont {Benenti}},\ }\bibfield  {title} {\enquote
  {\bibinfo {title} {Optimal energy conversion through antiadiabatic driving
  breaking time-reversal symmetry},}\ }\href {\doibase
  10.1103/PhysRevResearch.3.013237} {\bibfield  {journal} {\bibinfo  {journal}
  {Phys. Rev. Research}\ }\textbf {\bibinfo {volume} {3}},\ \bibinfo {pages}
  {013237} (\bibinfo {year} {2021})}\BibitemShut {NoStop}%
\bibitem [{\citenamefont {Esposito}\ \emph
  {et~al.}(2015{\natexlab{a}})\citenamefont {Esposito}, \citenamefont {Ochoa},\
  and\ \citenamefont {Galperin}}]{Esposito.15.PRL}%
  \BibitemOpen
  \bibfield  {author} {\bibinfo {author} {\bibfnamefont {M.}~\bibnamefont
  {Esposito}}, \bibinfo {author} {\bibfnamefont {M.~A.}\ \bibnamefont {Ochoa}},
  \ and\ \bibinfo {author} {\bibfnamefont {M.}~\bibnamefont {Galperin}},\
  }\bibfield  {title} {\enquote {\bibinfo {title} {Quantum thermodynamics: A
  nonequilibrium green's function approach},}\ }\href {\doibase
  10.1103/PhysRevLett.114.080602} {\bibfield  {journal} {\bibinfo  {journal}
  {Phys. Rev. Lett.}\ }\textbf {\bibinfo {volume} {114}},\ \bibinfo {pages}
  {080602} (\bibinfo {year} {2015}{\natexlab{a}})}\BibitemShut {NoStop}%
\bibitem [{\citenamefont {Esposito}\ \emph
  {et~al.}(2015{\natexlab{b}})\citenamefont {Esposito}, \citenamefont {Ochoa},\
  and\ \citenamefont {Galperin}}]{Esposito.15.PRB}%
  \BibitemOpen
  \bibfield  {author} {\bibinfo {author} {\bibfnamefont {M.}~\bibnamefont
  {Esposito}}, \bibinfo {author} {\bibfnamefont {M.~A.}\ \bibnamefont {Ochoa}},
  \ and\ \bibinfo {author} {\bibfnamefont {M.}~\bibnamefont {Galperin}},\
  }\bibfield  {title} {\enquote {\bibinfo {title} {Nature of heat in strongly
  coupled open quantum systems},}\ }\href {\doibase 10.1103/PhysRevB.92.235440}
  {\bibfield  {journal} {\bibinfo  {journal} {Phys. Rev. B}\ }\textbf {\bibinfo
  {volume} {92}},\ \bibinfo {pages} {235440} (\bibinfo {year}
  {2015}{\natexlab{b}})}\BibitemShut {NoStop}%
\bibitem [{\citenamefont {Bergmann}\ and\ \citenamefont
  {Galperin}(2021)}]{Bergmann.21.EPJ}%
  \BibitemOpen
  \bibfield  {author} {\bibinfo {author} {\bibfnamefont {N.}~\bibnamefont
  {Bergmann}}\ and\ \bibinfo {author} {\bibfnamefont {M.}~\bibnamefont
  {Galperin}},\ }\bibfield  {title} {\enquote {\bibinfo {title} {A green?s
  function perspective on the nonequilibrium thermodynamics of open quantum
  systems strongly coupled to baths},}\ }\href
  {https://doi.org/10.1140/epjs/s11734-021-00067-3} {\bibfield  {journal}
  {\bibinfo  {journal} {Eur. Phys. J. Spec. Top.}\ ,\ \bibinfo {pages} {--}}
  (\bibinfo {year} {2021})}\BibitemShut {NoStop}%
\bibitem [{Note1()}]{Note1}%
  \BibitemOpen
  \bibinfo {note} {One can obtain the present first law by combining the
  expression $\protect \mathcal {W}(m)=\DOTSI \intop \ilimits@ _{m\protect
  \mathcal {T}}^{(m+1)\protect \mathcal {T}}dt\protect \frac {d \delimiter
  "4360360 H(t)\delimiter "5365365 }{dt}$ with the Hamiltonian decomposition of
  Eq. (\ref {eq:H}).}\BibitemShut {Stop}%
\bibitem [{\citenamefont {Esposito}\ \emph {et~al.}(2010)\citenamefont
  {Esposito}, \citenamefont {Lindenberg},\ and\ \citenamefont {den
  Broeck}}]{Esposito.10.NJP}%
  \BibitemOpen
  \bibfield  {author} {\bibinfo {author} {\bibfnamefont {M.}~\bibnamefont
  {Esposito}}, \bibinfo {author} {\bibfnamefont {K.}~\bibnamefont
  {Lindenberg}}, \ and\ \bibinfo {author} {\bibfnamefont {C.~Van}\ \bibnamefont
  {den Broeck}},\ }\bibfield  {title} {\enquote {\bibinfo {title} {Entropy
  production as correlation between system and reservoir},}\ }\href {\doibase
  10.1088/1367-2630/12/1/013013} {\bibfield  {journal} {\bibinfo  {journal}
  {New J. Phys.}\ }\textbf {\bibinfo {volume} {12}},\ \bibinfo {pages} {013013}
  (\bibinfo {year} {2010})}\BibitemShut {NoStop}%
\bibitem [{\citenamefont {Ludovico}\ \emph {et~al.}(2014)\citenamefont
  {Ludovico}, \citenamefont {Lim}, \citenamefont {Moskalets}, \citenamefont
  {Arrachea},\ and\ \citenamefont {S\'anchez}}]{Ludovico.14.PRB}%
  \BibitemOpen
  \bibfield  {author} {\bibinfo {author} {\bibfnamefont {M.}~\bibnamefont
  {Ludovico}}, \bibinfo {author} {\bibfnamefont {J.}~\bibnamefont {Lim}},
  \bibinfo {author} {\bibfnamefont {M.}~\bibnamefont {Moskalets}}, \bibinfo
  {author} {\bibfnamefont {L.}~\bibnamefont {Arrachea}}, \ and\ \bibinfo
  {author} {\bibfnamefont {D.}~\bibnamefont {S\'anchez}},\ }\bibfield  {title}
  {\enquote {\bibinfo {title} {Dynamical energy transfer in ac-driven quantum
  systems},}\ }\href {\doibase 10.1103/PhysRevB.89.161306} {\bibfield
  {journal} {\bibinfo  {journal} {Phys. Rev. B}\ }\textbf {\bibinfo {volume}
  {89}},\ \bibinfo {pages} {161306} (\bibinfo {year} {2014})}\BibitemShut
  {NoStop}%
\bibitem [{\citenamefont {Ochoa}\ \emph {et~al.}(2016)\citenamefont {Ochoa},
  \citenamefont {Bruch},\ and\ \citenamefont {Nitzan}}]{Ochoa.16.PRB}%
  \BibitemOpen
  \bibfield  {author} {\bibinfo {author} {\bibfnamefont {M.~A.}\ \bibnamefont
  {Ochoa}}, \bibinfo {author} {\bibfnamefont {A.}~\bibnamefont {Bruch}}, \ and\
  \bibinfo {author} {\bibfnamefont {A.}~\bibnamefont {Nitzan}},\ }\bibfield
  {title} {\enquote {\bibinfo {title} {Energy distribution and local
  fluctuations in strongly coupled open quantum systems: The extended resonant
  level model},}\ }\href {\doibase 10.1103/PhysRevB.94.035420} {\bibfield
  {journal} {\bibinfo  {journal} {Phys. Rev. B}\ }\textbf {\bibinfo {volume}
  {94}},\ \bibinfo {pages} {035420} (\bibinfo {year} {2016})}\BibitemShut
  {NoStop}%
\bibitem [{\citenamefont {Thingna}\ \emph {et~al.}(2017)\citenamefont
  {Thingna}, \citenamefont {Barra},\ and\ \citenamefont
  {Esposito}}]{Thingna.17.PRE}%
  \BibitemOpen
  \bibfield  {author} {\bibinfo {author} {\bibfnamefont {J.}~\bibnamefont
  {Thingna}}, \bibinfo {author} {\bibfnamefont {F.}~\bibnamefont {Barra}}, \
  and\ \bibinfo {author} {\bibfnamefont {M.}~\bibnamefont {Esposito}},\
  }\bibfield  {title} {\enquote {\bibinfo {title} {Kinetics and thermodynamics
  of a driven open quantum system},}\ }\href {\doibase
  10.1103/PhysRevE.96.052132} {\bibfield  {journal} {\bibinfo  {journal} {Phys.
  Rev. E}\ }\textbf {\bibinfo {volume} {96}},\ \bibinfo {pages} {052132}
  (\bibinfo {year} {2017})}\BibitemShut {NoStop}%
\bibitem [{\citenamefont {Dou}\ \emph {et~al.}(2018)\citenamefont {Dou},
  \citenamefont {Ochoa}, \citenamefont {Nitzan},\ and\ \citenamefont
  {Subotnik}}]{Dou.18.PRB}%
  \BibitemOpen
  \bibfield  {author} {\bibinfo {author} {\bibfnamefont {W.}~\bibnamefont
  {Dou}}, \bibinfo {author} {\bibfnamefont {M.~A.}\ \bibnamefont {Ochoa}},
  \bibinfo {author} {\bibfnamefont {A.}~\bibnamefont {Nitzan}}, \ and\ \bibinfo
  {author} {\bibfnamefont {J.}~\bibnamefont {Subotnik}},\ }\bibfield  {title}
  {\enquote {\bibinfo {title} {Universal approach to quantum thermodynamics in
  the strong coupling regime},}\ }\href {\doibase 10.1103/PhysRevB.98.134306}
  {\bibfield  {journal} {\bibinfo  {journal} {Phys. Rev. B}\ }\textbf {\bibinfo
  {volume} {98}},\ \bibinfo {pages} {134306} (\bibinfo {year}
  {2018})}\BibitemShut {NoStop}%
\bibitem [{\citenamefont {Oz}\ \emph {et~al.}(2019)\citenamefont {Oz},
  \citenamefont {Hod},\ and\ \citenamefont {Nitzan}}]{Oz.19.MP}%
  \BibitemOpen
  \bibfield  {author} {\bibinfo {author} {\bibfnamefont {I.}~\bibnamefont
  {Oz}}, \bibinfo {author} {\bibfnamefont {O.}~\bibnamefont {Hod}}, \ and\
  \bibinfo {author} {\bibfnamefont {A.}~\bibnamefont {Nitzan}},\ }\bibfield
  {title} {\enquote {\bibinfo {title} {Evaluation of dynamical properties of
  open quantum systems using the driven liouville-von neumann approach:
  methodological considerations},}\ }\href {\doibase
  10.1080/00268976.2019.1584338} {\bibfield  {journal} {\bibinfo  {journal}
  {Mol. Phys.}\ }\textbf {\bibinfo {volume} {117}},\ \bibinfo {pages} {2083}
  (\bibinfo {year} {2019})}\BibitemShut {NoStop}%
\bibitem [{\citenamefont {Dou}\ \emph {et~al.}(2020)\citenamefont {Dou},
  \citenamefont {B\"atge}, \citenamefont {Levy},\ and\ \citenamefont
  {Thoss}}]{Dou.20.PRB}%
  \BibitemOpen
  \bibfield  {author} {\bibinfo {author} {\bibfnamefont {W.}~\bibnamefont
  {Dou}}, \bibinfo {author} {\bibfnamefont {J.}~\bibnamefont {B\"atge}},
  \bibinfo {author} {\bibfnamefont {A.}~\bibnamefont {Levy}}, \ and\ \bibinfo
  {author} {\bibfnamefont {M.}~\bibnamefont {Thoss}},\ }\bibfield  {title}
  {\enquote {\bibinfo {title} {Universal approach to quantum thermodynamics of
  strongly coupled systems under nonequilibrium conditions and external
  driving},}\ }\href {\doibase 10.1103/PhysRevB.101.184304} {\bibfield
  {journal} {\bibinfo  {journal} {Phys. Rev. B}\ }\textbf {\bibinfo {volume}
  {101}},\ \bibinfo {pages} {184304} (\bibinfo {year} {2020})}\BibitemShut
  {NoStop}%
\bibitem [{SM()}]{SM}%
  \BibitemOpen
  \href@noop {} {}\bibinfo {note} {See Supplemental Material for the estimation
  of heat-engine operation regime for the driven resonant level model,
  simulation details regarding the driven Liouville von-Neumann method and
  additional simulation results.}\BibitemShut {Stop}%
\bibitem [{\citenamefont {Liu}\ and\ \citenamefont {Segal}(2020)}]{LiuJ.20.NL}%
  \BibitemOpen
  \bibfield  {author} {\bibinfo {author} {\bibfnamefont {J.}~\bibnamefont
  {Liu}}\ and\ \bibinfo {author} {\bibfnamefont {D.}~\bibnamefont {Segal}},\
  }\bibfield  {title} {\enquote {\bibinfo {title} {Sharp negative differential
  resistance from vibrational mode softening in molecular junctions},}\ }\href
  {\doibase 10.1021/acs.nanolett.0c02230} {\bibfield  {journal} {\bibinfo
  {journal} {Nano Lett.}\ }\textbf {\bibinfo {volume} {20}},\ \bibinfo {pages}
  {6128} (\bibinfo {year} {2020})}\BibitemShut {NoStop}%
\bibitem [{\citenamefont {Ajisaka}\ \emph {et~al.}(2012)\citenamefont
  {Ajisaka}, \citenamefont {Barra}, \citenamefont {Mej\'{\i}a-Monasterio},\
  and\ \citenamefont {Prosen}}]{Ajisaka.12.PRB}%
  \BibitemOpen
  \bibfield  {author} {\bibinfo {author} {\bibfnamefont {S.}~\bibnamefont
  {Ajisaka}}, \bibinfo {author} {\bibfnamefont {F.}~\bibnamefont {Barra}},
  \bibinfo {author} {\bibfnamefont {C.}~\bibnamefont {Mej\'{\i}a-Monasterio}},
  \ and\ \bibinfo {author} {\bibfnamefont {T.}~\bibnamefont {Prosen}},\
  }\bibfield  {title} {\enquote {\bibinfo {title} {Nonequlibrium particle and
  energy currents in quantum chains connected to mesoscopic fermi
  reservoirs},}\ }\href {\doibase 10.1103/PhysRevB.86.125111} {\bibfield
  {journal} {\bibinfo  {journal} {Phys. Rev. B}\ }\textbf {\bibinfo {volume}
  {86}},\ \bibinfo {pages} {125111} (\bibinfo {year} {2012})}\BibitemShut
  {NoStop}%
\bibitem [{\citenamefont {Zelovich}\ \emph {et~al.}(2014)\citenamefont
  {Zelovich}, \citenamefont {Kronik},\ and\ \citenamefont
  {Hod}}]{Zelovich.14.JCTC}%
  \BibitemOpen
  \bibfield  {author} {\bibinfo {author} {\bibfnamefont {T.}~\bibnamefont
  {Zelovich}}, \bibinfo {author} {\bibfnamefont {L.}~\bibnamefont {Kronik}}, \
  and\ \bibinfo {author} {\bibfnamefont {O.}~\bibnamefont {Hod}},\ }\bibfield
  {title} {\enquote {\bibinfo {title} {State representation approach for
  atomistic time-dependent transport calculations in molecular junctions},}\
  }\href {\doibase 10.1021/ct500135e} {\bibfield  {journal} {\bibinfo
  {journal} {J. Chem. Theory Comput.}\ }\textbf {\bibinfo {volume} {10}},\
  \bibinfo {pages} {2927} (\bibinfo {year} {2014})}\BibitemShut {NoStop}%
\bibitem [{\citenamefont {Chen}\ \emph {et~al.}(2014)\citenamefont {Chen},
  \citenamefont {Hansen},\ and\ \citenamefont {Franco}}]{Chen.14.JPCC}%
  \BibitemOpen
  \bibfield  {author} {\bibinfo {author} {\bibfnamefont {L.}~\bibnamefont
  {Chen}}, \bibinfo {author} {\bibfnamefont {T.}~\bibnamefont {Hansen}}, \ and\
  \bibinfo {author} {\bibfnamefont {I.}~\bibnamefont {Franco}},\ }\bibfield
  {title} {\enquote {\bibinfo {title} {Simple and accurate method for
  time-dependent transport along nanoscale junctions},}\ }\href {\doibase
  10.1021/jp505771f} {\bibfield  {journal} {\bibinfo  {journal} {J. Phys. Chem.
  C}\ }\textbf {\bibinfo {volume} {118}},\ \bibinfo {pages} {20009} (\bibinfo
  {year} {2014})}\BibitemShut {NoStop}%
\bibitem [{\citenamefont {Zelovich}\ \emph {et~al.}(2017)\citenamefont
  {Zelovich}, \citenamefont {Hansen}, \citenamefont {Liu}, \citenamefont
  {Neaton}, \citenamefont {Kronik},\ and\ \citenamefont
  {Hod}}]{Zelovich.17.JCP}%
  \BibitemOpen
  \bibfield  {author} {\bibinfo {author} {\bibfnamefont {T.}~\bibnamefont
  {Zelovich}}, \bibinfo {author} {\bibfnamefont {T.}~\bibnamefont {Hansen}},
  \bibinfo {author} {\bibfnamefont {Z.}~\bibnamefont {Liu}}, \bibinfo {author}
  {\bibfnamefont {J.~B.}\ \bibnamefont {Neaton}}, \bibinfo {author}
  {\bibfnamefont {L.}~\bibnamefont {Kronik}}, \ and\ \bibinfo {author}
  {\bibfnamefont {O.}~\bibnamefont {Hod}},\ }\bibfield  {title} {\enquote
  {\bibinfo {title} {Parameter-free driven liouville-von neumann approach for
  time-dependent electronic transport simulations in open quantum systems},}\
  }\href {\doibase 10.1063/1.4976731} {\bibfield  {journal} {\bibinfo
  {journal} {J. Chem. Phys.}\ }\textbf {\bibinfo {volume} {146}},\ \bibinfo
  {pages} {092331} (\bibinfo {year} {2017})}\BibitemShut {NoStop}%
\bibitem [{\citenamefont {Chiang}\ and\ \citenamefont
  {Hsu}(2020)}]{Chiang.20.JCP}%
  \BibitemOpen
  \bibfield  {author} {\bibinfo {author} {\bibfnamefont {T-M.}\ \bibnamefont
  {Chiang}}\ and\ \bibinfo {author} {\bibfnamefont {L-Y.}\ \bibnamefont
  {Hsu}},\ }\bibfield  {title} {\enquote {\bibinfo {title} {Quantum transport
  with electronic relaxation in electrodes: Landauer-type formulas derived from
  the driven liouville?von neumann approach},}\ }\href {\doibase
  10.1063/5.0007750} {\bibfield  {journal} {\bibinfo  {journal} {J. Chem.
  Phys.}\ }\textbf {\bibinfo {volume} {153}},\ \bibinfo {pages} {044103}
  (\bibinfo {year} {2020})}\BibitemShut {NoStop}%
\bibitem [{\citenamefont {Newman}\ \emph {et~al.}(2020)\citenamefont {Newman},
  \citenamefont {Mintert},\ and\ \citenamefont {Nazir}}]{Newman.20.PRE}%
  \BibitemOpen
  \bibfield  {author} {\bibinfo {author} {\bibfnamefont {D.}~\bibnamefont
  {Newman}}, \bibinfo {author} {\bibfnamefont {F.}~\bibnamefont {Mintert}}, \
  and\ \bibinfo {author} {\bibfnamefont {A.}~\bibnamefont {Nazir}},\ }\bibfield
   {title} {\enquote {\bibinfo {title} {Quantum limit to nonequilibrium
  heat-engine performance imposed by strong system-reservoir coupling},}\
  }\href {\doibase 10.1103/PhysRevE.101.052129} {\bibfield  {journal} {\bibinfo
   {journal} {Phys. Rev. E}\ }\textbf {\bibinfo {volume} {101}},\ \bibinfo
  {pages} {052129} (\bibinfo {year} {2020})}\BibitemShut {NoStop}%
\end{thebibliography}

%

\newpage
\renewcommand{\thesection}{\Roman{section}} 
\renewcommand{\thesubsection}{\Alph{subsection}}
\renewcommand{\theequation}{S\arabic{equation}}
\renewcommand{\thefigure}{S\arabic{figure}}
\renewcommand{\thetable}{S\arabic{table}}
\setcounter{equation}{0}  
\setcounter{figure}{0}

\begin{widetext}

{\Large{\bf Supplemental material:} Periodically-driven quantum thermal machines from warming up to limit cycle}
\\
\\
\\

In this supplemental material, we first identify the parameter regime where the driven resonant level model operates as a quantum Otto heat engine.
Next, we provide simulation details regarding the numerical implementation of the driven Liouville von-Neumann (DLvN) approach.
We also present additional simulation results that complement those included in the main text.

\section{I. Regime of heat engine behavior of the driven resonant level model}
Here we determine the operation regime of a quantum Otto heat engine made of a driven resonant level model from a general perspective.
We rely on the facts that (i) only strokes 2 and 4 of a quantum Otto cycle produce nonzero work and (ii) that the
charge occupation of the central level remains unaltered during these two strokes. 
We assume here that the engine has reached the limit-cycle phase at $t=0$, starting from the remote past $t\to-\infty$, such that we can resort to the integrated injected work obtained within a cycle $[0,\mathcal{T}]$ to access the sign of work, 
\bea
\mathcal{W} &=& \int_0^{\mathcal{T}}dt~\dot{\epsilon}(t)\langle n_d(t)\rangle\nonumber\\
&=& \int_{t_1}^{t_1+t_2}dt~\dot{\epsilon}(t)\langle n_d(t)\rangle+\int_{t_1+t_2+t_3}^{\mathcal{T}}dt~\dot{\epsilon}(t)\langle n_d(t)\rangle.
\eea
Here, $\langle n_d(t)\rangle=\langle c_d^{\dagger}(t)c_d(t)\rangle$. 
Since $\langle n_d\rangle$ remains fixed during strokes 2 and 4, we simplify the above expression as
\bea\label{eq:w_esti}
\mathcal{W} &=& \langle n_d\rangle_h\int_{t_1}^{t_1+t_2}dt~\dot{\epsilon}(t)+\langle n_d\rangle_c \int_{t_1+t_2+t_3}^{\mathcal{T}}dt~\dot{\epsilon}(t)\nonumber\\
&=& \langle n_d\rangle_h (\epsilon_2-\epsilon_1)+\langle n_d\rangle_c(\epsilon_1-\epsilon_2).
\eea
\begin{figure}[tbh!]
 \centering
\includegraphics[width=0.8\columnwidth] {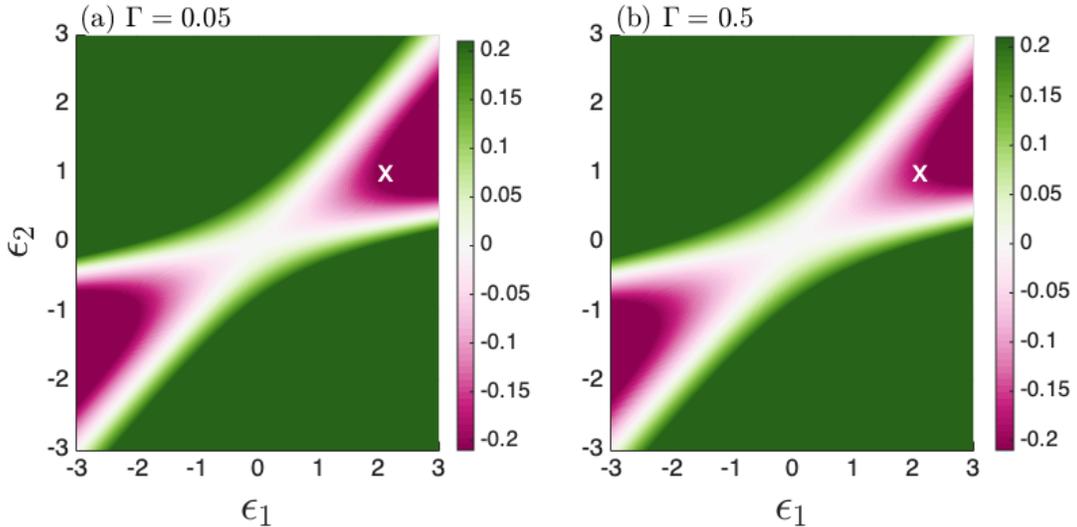}
\caption{Injected work during a limit cycle, $\mathcal{W}$, as a function of $\epsilon_{1,2}$ obtained using Eqs. (\ref{eq:w_esti}) and (\ref{eq:nd}): (a) $\Gamma_h=\Gamma_c=0.05$, (b) $\Gamma_h=\Gamma_c=0.5$.
White crosses mark the heat engine conditions $\epsilon_1=2$ and $\epsilon_2=1$ used throughout the present study. 
Other dimensionless parameters are $\beta_h=0.2$, $\beta_c=1.5$ and $\mu_h=\mu_c=0$.
}
\label{fig:work}
\end{figure}
Here, $\langle n_d\rangle_{h(c)}$ denote the occupation after the hot (cold) isochore. We have used the fact that $\epsilon(t_1)=\epsilon(\mathcal{T})=\epsilon_1$, $\epsilon(t_1+t_2)=\epsilon(t_1+t_2+t_3)=\epsilon_2$, which stems from the definition of our quantum Otto cycle (see Fig. 1 of the main text).

Since we can set the time intervals $t_{1,3}$ long enough to allow complete equilibration, 
$\langle n_d\rangle_{h(c)}$ can be estimated using their equilibrium expressions (see, e.g., Ref. \cite{LiuJ.20.NL})
\bea\label{eq:nd}
\langle n_d\rangle_{h} &=& \Gamma\int\,\frac{d\epsilon}{2\pi}\frac{n_F^h(\epsilon)}{(\Gamma/2)^2+(\epsilon-\epsilon_1)^2},\nonumber\\
\langle n_d\rangle_{c} &=& \Gamma\int\,\frac{d\epsilon}{2\pi}\frac{n_F^c(\epsilon)}{(\Gamma/2)^2+(\epsilon-\epsilon_2)^2}.
\eea
Here, we considered symmetric coupling with $\Gamma_h=\Gamma_c=\Gamma$, $n_F^v(\epsilon)=[\exp(\beta_v(\epsilon-\mu_v))+1]^{-1}$ is the Fermi-Dirac distribution of the $v$-lead characterized by an inverse temperature $\beta_v$ and a chemical potential $\mu_v$. By combining Eqs. (\ref{eq:w_esti}) and (\ref{eq:nd}), we determine the operation regimes of a heat engine, i.e. for which parameters the work per cycle, in the limit cycle region is negative, $\mathcal{W}<0$. 
Fig. \ref{fig:work} depicts a set of calculation results while varying $\epsilon_{1,2}$ for two different coupling strengths $\Gamma$; heat engine operation regimes with negative $\mathcal{W}$ are highlighted by the red zones. For simulation purposes, throughout this study  we choose  the values $\epsilon_1=2$ and $\epsilon_2=1$ as marked by white crosses in the figure.

\section{II. Driven Liouville von-Neumann approach}
We employ the so-called driven Liouville von-Neumann (DLvN) approach \cite{Oz.19.MP,Oz.20.JCTC,Chiang.20.JCP} to simulate the dynamics of the driven resonant level model. Since the total Hamiltonian of the driven resonant level model is quadratic, all observables can be constructed from the correlator $\sigma_{ij}(t)=\langle c_i^{\dagger}(t)c_j(t)\rangle$ with $i,j=d,~kv$. The average is evaluated with respect to the total initial state. For simplicity, we denote $\sigma_i=\sigma_{ii}$ hereafter. To calculate the dynamics of thermodynamic observables, we introduce the following matrix
\begin{equation}
\boldsymbol{\sigma}~=~\left(\begin{array}{ccc}
\boldsymbol{\sigma}_L & \boldsymbol{\sigma}_{Ld} & \boldsymbol{0}\\
\boldsymbol{\sigma}_{dL} & \sigma_d & \boldsymbol{\sigma}_{dR}\\
\boldsymbol{0} & \boldsymbol{\sigma}_{Rd} & \boldsymbol{\sigma}_R
\end{array}
\right).
\end{equation}
Here, $\sigma_d$ stands for the dot population, $\boldsymbol{\sigma}_{L,R}$ are diagonal blocks of leads, and $\boldsymbol{\sigma}_{dv}=\boldsymbol{\sigma}_{vd}^{\dagger}$ are the dot-lead coherences vectors. Correspondingly, the total Hamiltonian matrix is expressed as
\begin{equation}
\boldsymbol{H}(t)~=~\left(\begin{array}{ccc}
\boldsymbol{\epsilon}_L & \lambda_L(t)\boldsymbol{t}_L & \boldsymbol{0}\\
 \lambda_L(t)\boldsymbol{t}_L ^{\dagger}& \epsilon_d(t) & \lambda_R(t)\boldsymbol{t}_R^{\dagger}\\
\boldsymbol{0} & \lambda_R(t)\boldsymbol{t}_R & \boldsymbol{\epsilon}_R
\end{array}
\right).
\end{equation}
Here, $\boldsymbol{\epsilon}_{L,R}$ are diagonal blocks of lead level energies, $\epsilon_{kv}$, which are taken to be equally spaced over the bandwidth in numerical simulations.  $\boldsymbol{t}_{L,R}$ are column vectors containing tunneling elements $t_{kv}$. With the above definitions, the driven Liouville von-Neumann (DLvN) equation governing the evolution of $\boldsymbol{\sigma}(t)$ takes the following form \cite{Oz.19.MP,Oz.20.JCTC,Chiang.20.JCP}
\begin{equation}\label{eq:lvn}
\frac{d}{dt}\boldsymbol{\sigma}(t)~=~-i[\boldsymbol{H}(t),\boldsymbol{\sigma}(t)]-\gamma\left(\begin{array}{ccc}
\boldsymbol{\sigma}_L(t)-\boldsymbol{\sigma}_L^{\mathrm{eq}} & \frac{1}{2}\boldsymbol{\sigma}_{Ld}(t) & \boldsymbol{0}\\
\frac{1}{2}\boldsymbol{\sigma}_{dL}(t) & 0 & \frac{1}{2}\boldsymbol{\sigma}_{dR}(t)\\
\boldsymbol{0} & \frac{1}{2}\boldsymbol{\sigma}_{Rd}(t) & \boldsymbol{\sigma}_R(t)-\boldsymbol{\sigma}_R^{\mathrm{eq}}
\end{array}\right).
\end{equation}
Here, the equilibrium correlator elements are $[\boldsymbol{\sigma}_{L,R}^{\mathrm{eq}}]_{k,k'}=\delta_{kk'}n_F^{L,R}(\epsilon_{kv})$. The lead driving
 rate constant $\gamma$ is set to be of the order of the lead energy spacing. It represents the timescale on which thermal relaxation takes place in the leads due to their couplings to an implicit superbath \cite{Oz.19.MP,Oz.20.JCTC}. 
 We check and show below in Figs. \ref{fig:converg}-\ref{fig:convergS} that the dynamics of thermodynamic observables is fairly insensitive to the specific choice 
 of $\gamma$ in the coupling range for $\Gamma$ considered in this study. 

\subsection{A. Numerical implementation}
To simplify the implementation of the DLvN approach, we consider the wideband limit for the leads, $\Gamma_v(\epsilon)=\Gamma_v$ such that $t_{kv}=t_v$ with $t_v$ determined by $\Gamma_v=2\pi t_v^2\rho_v$ with $\rho_v$ a constant density of state of the $v$-lead \cite{Oz.19.MP,Oz.20.JCTC}. In numerical treatment, we represent each lead by a single energy band of a finite bandwidth $2D_v$ (specifically, electronic levels in the metals extend from $-D_v$ to $D_v$). We use a finite number, $N_v$, of equally spaced single-electron states with energy spacing $\Delta\epsilon_v=2D_v/N_v$ ($\rho_v=\Delta\epsilon_v^{-1}$ accordingly). Note that to get reasonable results, we require that $\Gamma_v>\Delta\epsilon_v$ and $2D_v>\Gamma_v$. The DLvN equation Eq. (\ref{eq:lvn}) is subject to an initial factorized state
\begin{equation}
\boldsymbol{\sigma}_0~=~\left(\begin{array}{ccc}
\boldsymbol{\sigma}_L^{\mathrm{eq}} & \boldsymbol{0} & \boldsymbol{0}\\
\boldsymbol{0} & 0 & \boldsymbol{0}\\
\boldsymbol{0} & \boldsymbol{0} & \boldsymbol{\sigma}_R^{\mathrm{eq}}.
\end{array}
\right).
\end{equation}
It is then numerically evolved using the fourth-order Runge-Kutta method with a fixed time step $\delta t$. 
We check the convergence of results with respect to $\Delta\epsilon_v$, $D_v$ and $\delta t$ (see below Figs. \ref{fig:converg}-\ref{fig:convergS} for examples). 

\subsection{B. Working expressions for thermodynamic observables}
In terms of the correlator $\boldsymbol{\sigma}(t)$ and Hamiltonian $\boldsymbol{H}(t)$, the work and $\mathcal{A}$ term can be directly obtained as
\bea
\mathcal{W}(m) &=& \int_{m\mathcal{T}}^{(m+1)\mathcal{T}}dt\mathrm{Tr}[\boldsymbol{\sigma}(t)\dot{\boldsymbol{H}}(t)],\nonumber\\
\mathcal{A}(m) &=& \int_{m\mathcal{T}}^{(m+1)\mathcal{T}}dt\frac{\partial}{\partial t}\mathrm{Tr}\Big\{\left[\boldsymbol{H}_S(t)+\boldsymbol{H}_I(t)\right]\boldsymbol{\sigma}(t)\Big\}.
\eea
Here, $\dot{O}\equiv dO/dt$. We also introduced the following matrix forms
\bea
\boldsymbol{H}_S(t) &=& \left(\begin{array}{ccc}
\boldsymbol{0} & \boldsymbol{0} & \boldsymbol{0}\\
\boldsymbol{0} & \epsilon_d(t) & \boldsymbol{0}\\
\boldsymbol{0} & \boldsymbol{0} & \boldsymbol{0}
\end{array}
\right),\nonumber\\
\boldsymbol{H}_I(t) &=& \left(\begin{array}{ccc}
\boldsymbol{0} & \lambda_L(t)\boldsymbol{t}_L & \boldsymbol{0}\\
 \lambda_L(t)\boldsymbol{t}_L^{\dagger} & 0 & \lambda_R(t)\boldsymbol{t}_R^{\dagger}\\
\boldsymbol{0} & \lambda_R(t)\boldsymbol{t}_R & \boldsymbol{0}
\end{array}
\right).
\eea
The heat $\mathcal{Q}_v$ absorbed by the resonant level from the $v$ bath should be evaluated as
\begin{equation}
\mathcal{Q}_v(m) ~=~ -\int_{m\mathcal{T}}^{(m+1)\mathcal{T}}dt\mathrm{Tr}[\boldsymbol{H}_B^v\dot{\boldsymbol{\sigma}}(t)]-\gamma\int_{m\mathcal{T}}^{(m+1)\mathcal{T}}dt\mathrm{Tr}[\boldsymbol{Z}_v\boldsymbol{H}(t)],
\end{equation}
where the reservoir matrices read
\bea
\boldsymbol{H}_B^L &=& \left(\begin{array}{ccc}
\boldsymbol{\epsilon}_L & \boldsymbol{0} & \boldsymbol{0}\\
\boldsymbol{0} & 0 & \boldsymbol{0}\\
\boldsymbol{0} & \boldsymbol{0} & \boldsymbol{0}
\end{array}
\right),\nonumber\\
\boldsymbol{H}_B^R &=& \left(\begin{array}{ccc}
\boldsymbol{0} & \boldsymbol{0} & \boldsymbol{0}\\
\boldsymbol{0} & 0 & \boldsymbol{0}\\
\boldsymbol{0} & \boldsymbol{0} & \boldsymbol{\epsilon}_R
\end{array}
\right),
\eea
and the second term on the right-hand-side of Eq. (\ref{eq:heat}) denotes the heat current exchange between the $v$-lead and its implicit bath (see Eq. (59) of Ref. \cite{Oz.20.JCTC}) with
\bea
\boldsymbol{Z}_L &=& \left(\begin{array}{ccc}
\boldsymbol{\sigma}_L-\boldsymbol{\sigma}_L^{\mathrm{eq}} & \frac{1}{2}\boldsymbol{\sigma}_{Ld} & \boldsymbol{0}\\
\frac{1}{2}\boldsymbol{\sigma}_{dL} & 0 & \boldsymbol{0}\\
\boldsymbol{0} & \boldsymbol{0} & \boldsymbol{0}
\end{array}\right),\nonumber\\
\boldsymbol{Z}_R &=& \left(\begin{array}{ccc}
\boldsymbol{0} & \boldsymbol{0} & \boldsymbol{0}\\
\boldsymbol{0} & 0 & \frac{1}{2}\boldsymbol{\sigma}_{dR}\\
\boldsymbol{0} & \frac{1}{2}\boldsymbol{\sigma}_{Rd} & \boldsymbol{\sigma}_R-\boldsymbol{\sigma}_R^{\mathrm{eq}}
\end{array}\right).
\eea
Using the DLvN equation (\ref{eq:lvn}) and noting that $\mathrm{Tr}[\boldsymbol{H}_B^v\dot{\boldsymbol{\sigma}}(t)]=\frac{d}{dt}\mathrm{Tr}[\boldsymbol{H}_B^v\boldsymbol{\sigma}(t)]$ and $\mathrm{Tr}\big\{\boldsymbol{H}(t)[\boldsymbol{H}(t),\boldsymbol{\sigma}(t)]\big\}=0$, we readily check that the first law of thermodynamics,
\begin{equation}
\mathcal{W}(m)+\sum_v\mathcal{Q}_v(m)-\mathcal{A}(m)~=~0,
\end{equation}
is satisfied by the above expressions.

\section{III. Driving protocol and convergence check}
To facilitate detailed numerical simulations, we need to specify the functional form of $\epsilon(t)$. 
The overall work output within a cycle is independent of the detailed form of $\epsilon(t)$, as we explained in Sec. I.
Therefore, we have the freedom to choose its functional form with the understanding that it should be at least once differentiable (since $\dot{\epsilon}(t)$ is needed so as to determine the amount of generated work during a cycle) and $\epsilon(t) = \epsilon(t+\mathcal{T})$. Specifically, we adopt the following protocol for $\epsilon(t)$ within a period $[0,\mathcal{T}]$ (see Fig. \ref{fig:ept}):
\begin{equation}
\epsilon(t) = \begin{cases}
\epsilon_1 & \text{for $0 \leq t < t_1$; stroke 1}\\
\epsilon_1-(\epsilon_1-\epsilon_2)\cdot Z(o_1) & \text{for $t_1 \leq t < t_1+t_2$; stroke 2}\\
\epsilon_2 & \text{for $t_1+t_2 \leq t < t_1+t_2+t_3$; stroke 3}\\
\epsilon_2+ (\epsilon_1-\epsilon_2)\cdot Z(o_2) & \text{for $t_1+t_2+t_3 \leq t < \mathcal{T}$; stroke 4}.
\end{cases}
\label{eq:om_t}
\end{equation}
where $Z(s) = 3s^2 - 2s^3$, $o_1=(t-t_1)/t_2$ and $o_2=(t-t_1-t_2-t_3)/t_4$. 
\begin{figure}[h]
\centering
\includegraphics[width=0.55\textwidth]{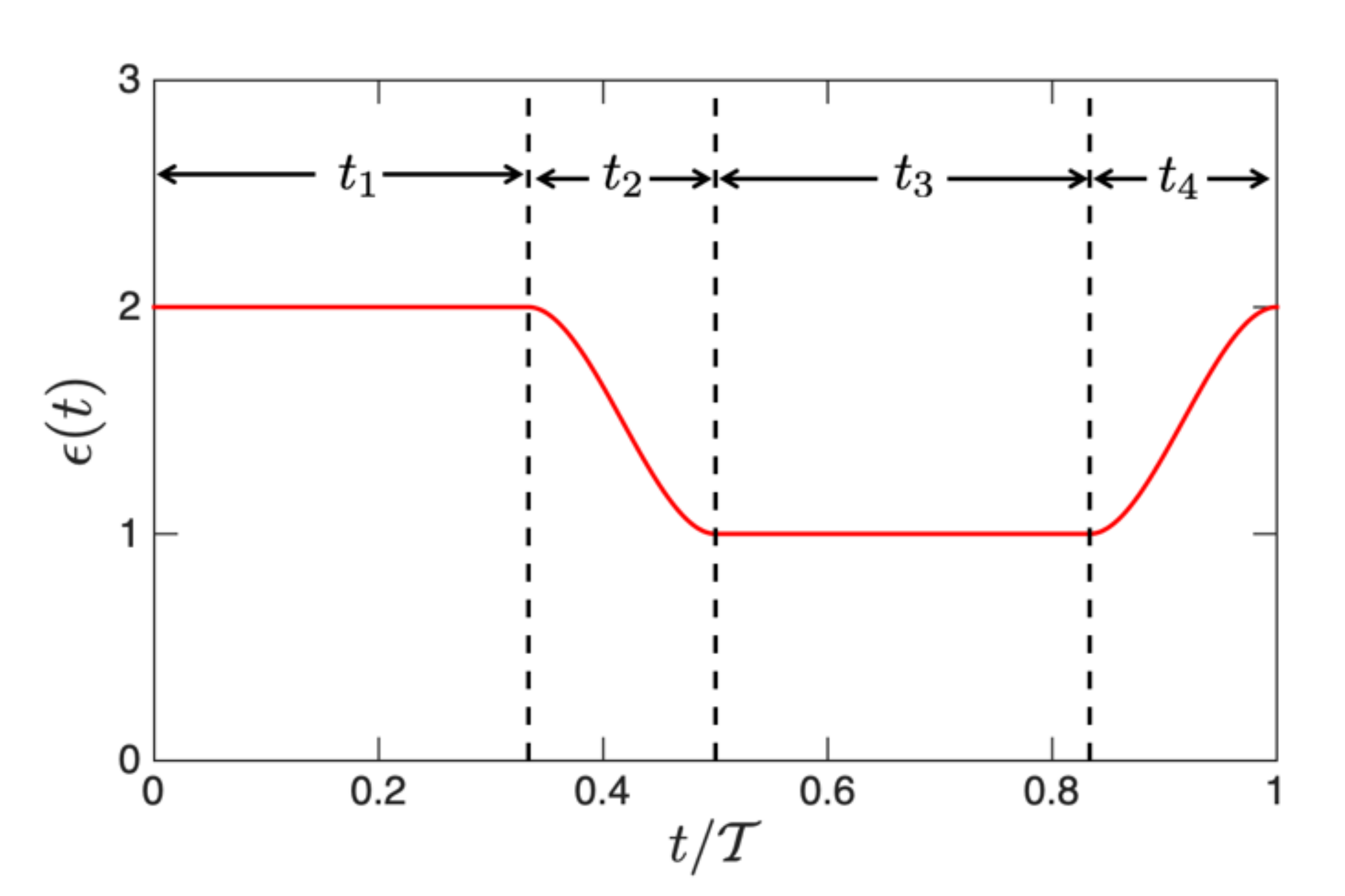} 
\caption{
Functional form of $\epsilon(t)$ (red solid line) determined by Eq. (\ref{eq:om_t}) with $\epsilon_1=2$, $\epsilon_2=1$, $t_1=t_3=\mathcal{T}/3$ and $t_2=t_4=\mathcal{T}/6$.
}
\protect\label{fig:ept}
\end{figure}
\begin{figure}[htbp]
\centering
\includegraphics[width=0.7\textwidth]{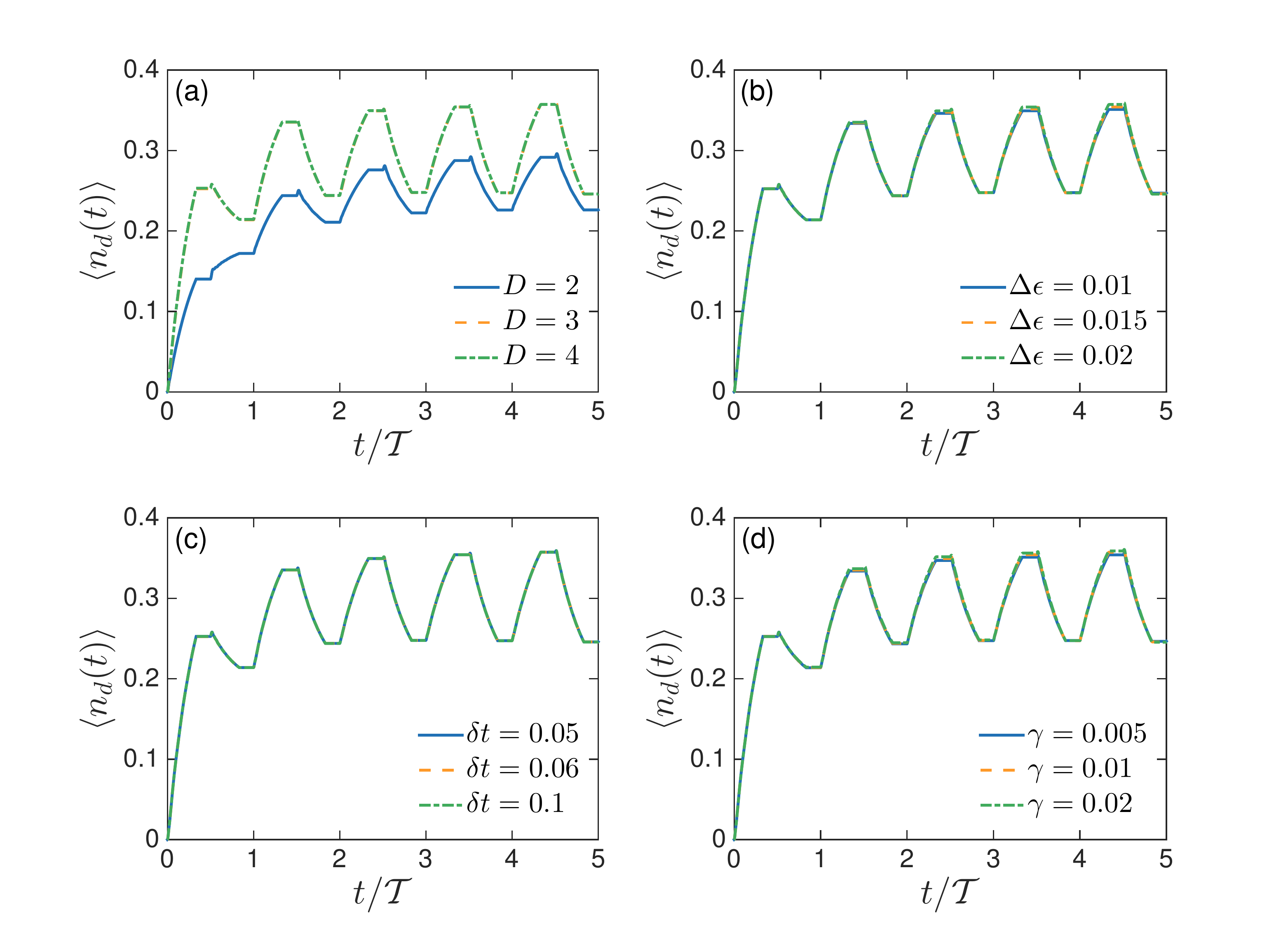} 
\caption{Convergence analysis at weak coupling. 
Charge occupation $\langle n_d(t)\rangle$ as a function of time while varying: (a) the bandwidth $D$ at fixed $\Delta\epsilon=0.02$, $\gamma=0.01$ and $\delta t=0.05$; results at large $D$ overlap.
 (b)  lead energy spacing $\Delta\epsilon$ (identical for the left and right metals). When the spacing is fine, it does not affect the occupation. 
 We  used $D=3$, $\gamma=0.01$ and $\delta t=0.05$.
  (c) time step $\delta t$ with fixed $D=3$, $\gamma=0.01$, $\Delta\epsilon=0.02$ and (d) damping rate constant $\gamma$ with fixed $D=3$, $\Delta\epsilon=0.02$ and $\delta t=0.05$. Other dimensionless parameters are $\beta_h=0.2$, $\beta_c=1.5$, $\mu_h=\mu_c=0$, $\epsilon_1=2$, $\epsilon_2=1$, $\Gamma=0.05$, $t_1=t_3=\mathcal{T}/3$ and $t_2=t_4=\mathcal{T}/6$ with $\mathcal{T}=60$.
}
\protect\label{fig:converg}
\end{figure}
\begin{figure}[htbp]
\centering
\includegraphics[width=0.7\textwidth]{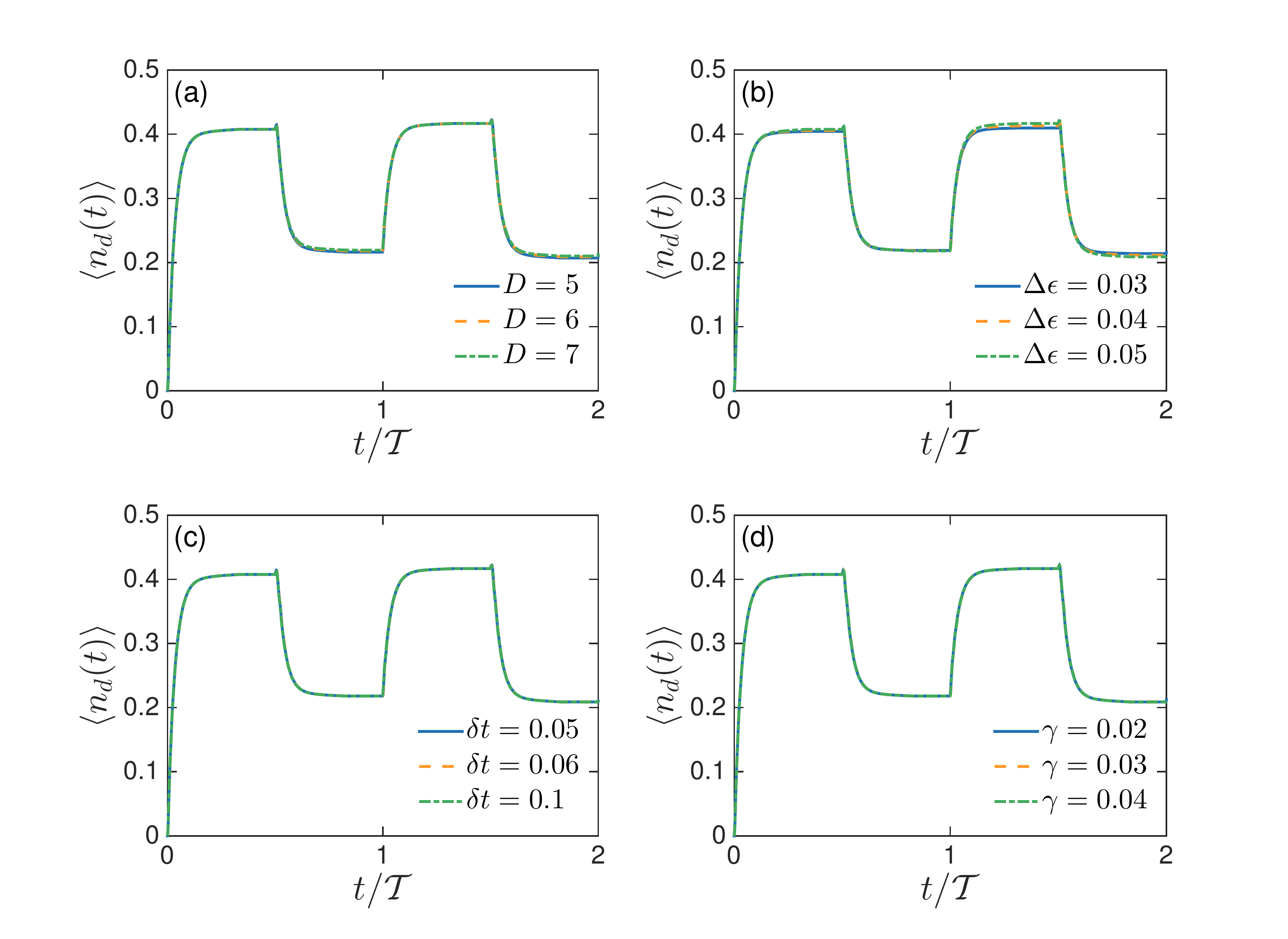} 
\caption{Convergence analysis at strong coupling. 
Charge occupation $\langle n_d(t)\rangle$ as a function of time while varying: (a) the bandwidth $D$ at fixed $\Delta\epsilon=0.02$, $\gamma=0.01$ and $\delta t=0.05$; results at large $D$ overlap.
 (b)  lead energy spacing $\Delta\epsilon$ (identical for the left and right metals). When the spacing is fine, it does not affect the occupation. 
 We  used $D=3$, $\gamma=0.01$ and $\delta t=0.05$.
  (c) time step $\delta t$ with fixed $D=3$, $\gamma=0.01$, $\Delta\epsilon=0.02$ and (d) damping rate constant $\gamma$ with fixed $D=3$, $\Delta\epsilon=0.02$ and $\delta t=0.05$. Other dimensionless parameters are $\beta_h=0.2$, $\beta_c=1.5$, $\mu_h=\mu_c=0$, $\epsilon_1=2$, $\epsilon_2=1$, $\Gamma=0.5$, $t_1=t_3=\mathcal{T}/3$ and $t_2=t_4=\mathcal{T}/6$ with $\mathcal{T}=60$.
}
\protect\label{fig:convergS}
\end{figure}


Using the above driving protocol [cf, Eq. (\ref{eq:om_t})] within the DLvN equation [cf, Eq. (\ref{eq:lvn})], we readily check the convergence of various quantities with respect to phenomenological damping rate $\gamma$, band width $D_v=D$, lead energy spacing $\Delta\epsilon_v=\Delta\epsilon$ and evolution time step $\delta t$. Considering the weak coupling limit, in Fig. \ref{fig:converg}, we show a typical set of results for charge occupation $\langle n_d(t)\rangle$ of the resonant level when varying $D$ [panel (a)], $\Delta\epsilon$ [panel (b)], $\delta t$ [panel (c)] and $\gamma$ [panel (d)].
Analogous simulations are presented in Fig. \ref{fig:convergS}, using $\Gamma=0.5$. 

We summarize in Table I parameter sets used in the present study, which lead to well-converged results.
\begin{table}[tbh!]
\caption{Converged parameter sets with varying level-lead coupling strength $\Gamma_v$}
\centering
\begin{tabularx}{0.6\textwidth} { 
  | >{\centering\arraybackslash}X
  | >{\centering\arraybackslash}X
  | >{\centering\arraybackslash}X
  | >{\centering\arraybackslash}X 
  | >{\centering\arraybackslash}X | }
 \hline
 & $\Gamma=0.02$ & $\Gamma=0.05$ & $\Gamma=0.2$ & $\Gamma=0.5$ \\
 \hline
$D$  & 3 & 3 & 5 & 6  \\
 \hline
$\Delta\epsilon$  & 0.006 & 0.015 & 0.0125 & 0.03  \\
 \hline
$\delta t$  & 0.1 & 0.1 & 0.1 & 0.1  \\
 \hline
$\gamma$  & 0.006  & 0.015 & 0.0125 & 0.03  \\
\hline
\end{tabularx}
\label{tab:1}
\end{table}

\newpage
\section{IV. Additional simulation results}
Here we present thermodynamic results that complement the results in the main text.

\subsection{A. Heat}
In Fig. \ref{fig:heat}, we present results for heat exchange with increasing cycle number $m$.
\begin{figure}[h]
\centering
\includegraphics[width=0.7\textwidth]{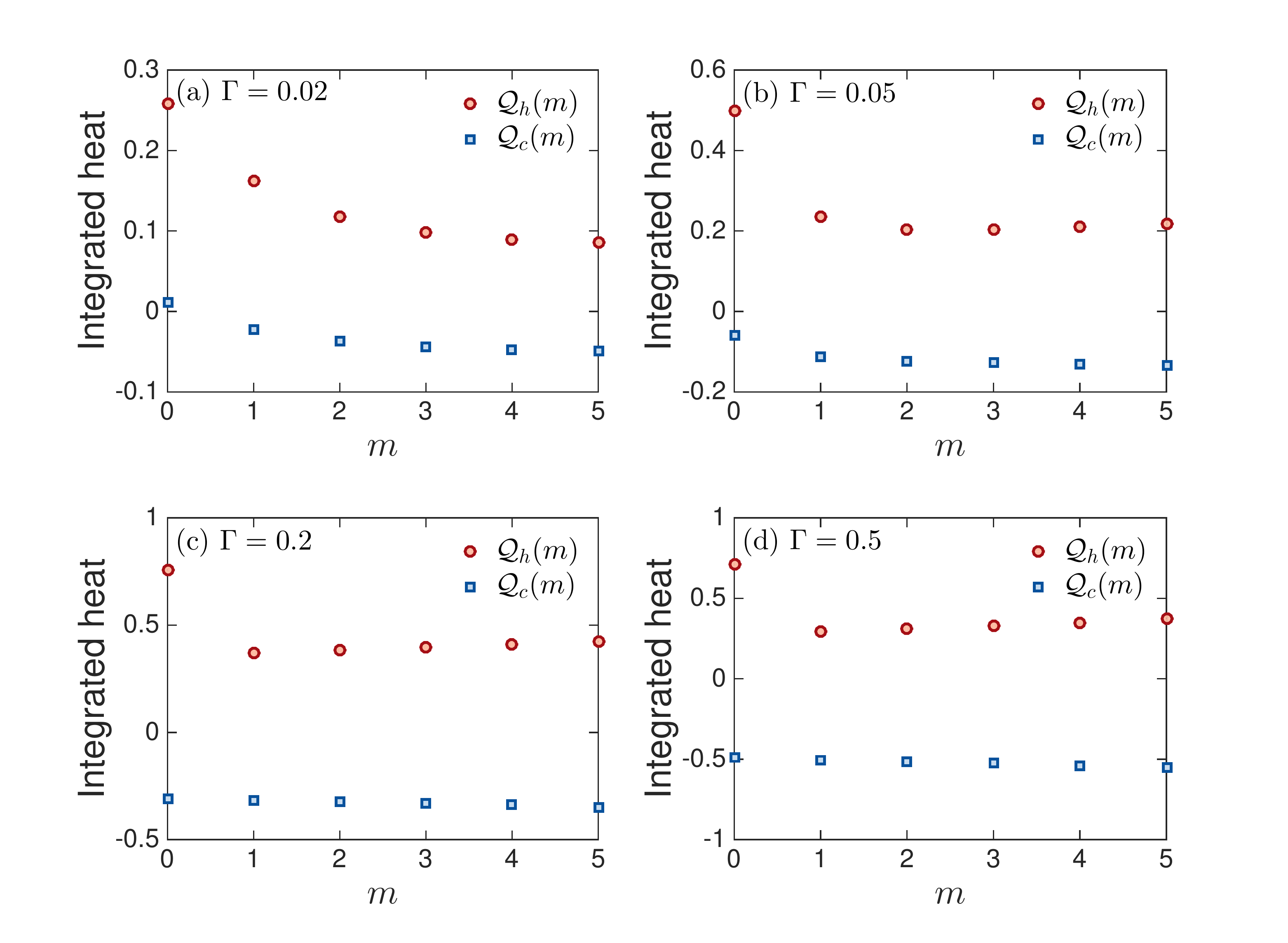}
\caption{Integrated heats $\mathcal{Q}_{h}(m)$ (red circles) and $\mathcal{Q}_c(t)$ (blue squares) within a cycle $[m\mathcal{T},(m+1)\mathcal{T}]$ with increasing $m$ (note that $m+1$ is the number of completed cycles) for (a) $\Gamma=0.02$, (b) $\Gamma=0.05$, (c) $\Gamma=0.2$ and (d) $\Gamma=0.5$. Other dimensionless parameters are $\beta_h=0.2$, $\beta_c=1.5$, $\mu_h=\mu_c=0$, $\epsilon_1=2$, $\epsilon_2=1$, $t_1=t_3=\mathcal{T}/3$ and $t_2=t_4=\mathcal{T}/6$ with $\mathcal{T}=60$.
}
\protect\label{fig:heat}
\end{figure}
Two points are to be highlighted, illustrating the important aspects of the present study, namely that
periodicity can break down at the ensemble average level and the presence of $\mathcal{A}$ term is a necessity: 

(i) In the limit-cycle phase as marked by stationary $\mathcal{A}$ term (see discussion in the main text regarding this point), the heats $\mathcal{Q}_{c,h}(m)$ still vary as the cycle number $m$ increases. 
This phenomenon is particularly clear at strong couplings (see Fig. \ref{fig:heat} (c) (d)). Since heat is defined as the change of the bath (lead) energy within a cycle,  it thus depends only on the lead density matrix. One can therefore infer that the lead density matrix needs not inherit the periodicity of the system density matrix, as argued in the main text.   
(ii) At large coupling strength, we observe that $|\mathcal{Q}_c|>|\mathcal{Q}_h|$ (Fig. \ref{fig:heat} (d)). We remark that this seemingly unphysical result is allowed by the first law of thermodynamics as this imbalance is precisely compensated by a negative $\mathcal{A}$ term (see Fig. 2 (b) of the main text).

\subsection{B. Long-time behavior at weak coupling}
In Fig. \ref{fig:long}, we illustrate the long-time behavior for a weak-coupling scenario, $\Gamma=0.02$.
%
\begin{figure}[h]
\centering
\includegraphics[width=0.7\textwidth]{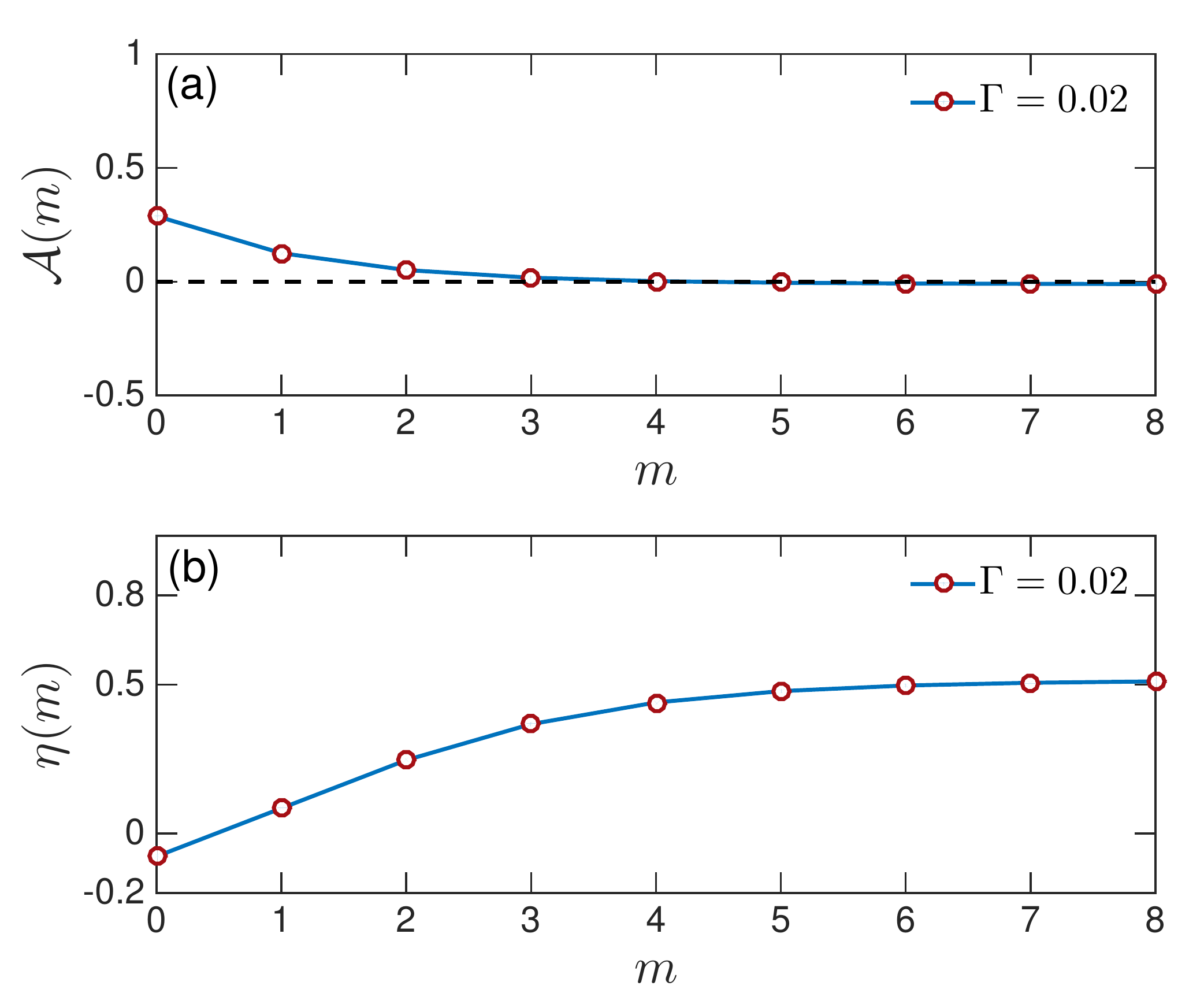} 
\caption{(a) $\mathcal{A}(m)$ with increasing $m$ (note that $m+1$ is the number of completed cycles) for $\Gamma=0.02$. (b) Thermodynamic efficiency $\eta(m)$ as a function of $m$ for $\Gamma=0.02$. Other dimensionless parameters are $\beta_h=0.2$, $\beta_c=1.5$, $\mu_h=\mu_c=0$, $\epsilon_1=2$, $\epsilon_2=1$, $t_1=t_3=\mathcal{T}/3$ and $t_2=t_4=\mathcal{T}/6$ with $\mathcal{T}=60$.
}
\protect\label{fig:long}
\end{figure}
From the figure, we confirm that the system indeed reaches its limit-cycle phase at $m\approx5$, as argued in the main text.

\end{widetext}

\end{document}